\documentclass[reprint,superscriptaddress,amsmath,amssymb,aps]{revtex4-2}

\usepackage{graphicx}
\usepackage{dcolumn}
\usepackage{booktabs, floatrow, makecell}
\usepackage{float}
\usepackage{bm}
\usepackage{lipsum}
\usepackage{xcolor} 
\usepackage{xfrac}
\usepackage{siunitx}
\usepackage{gensymb}

\usepackage{soul} 
\usepackage{scalerel}
\usepackage{tikz}
\usepackage[normalem]{ulem} 
\usepackage{tabularx} 

\usepackage{hyperref}
\hypersetup{colorlinks=true,citecolor=blue,linkcolor=blue,urlcolor=blue}

\usetikzlibrary{svg.path}

\usepackage{catchfile}

\newcommand{\comment}[1]{\textcolor{red}{#1}}

\newcommand{\norm}[1]{\left\|#1\right\|}

\definecolor{orcidlogocol}{HTML}{A6CE39}
\tikzset{
  orcidlogo/.pic={
    \fill[orcidlogocol] svg{M256,128c0,70.7-57.3,128-128,128C57.3,256,0,198.7,0,128C0,57.3,57.3,0,128,0C198.7,0,256,57.3,256,128z};
    \fill[white] svg{M86.3,186.2H70.9V79.1h15.4v48.4V186.2z}
                 svg{M108.9,79.1h41.6c39.6,0,57,28.3,57,53.6c0,27.5-21.5,53.6-56.8,53.6h-41.8V79.1z M124.3,172.4h24.5c34.9,0,42.9-26.5,42.9-39.7c0-21.5-13.7-39.7-43.7-39.7h-23.7V172.4z}
                 svg{M88.7,56.8c0,5.5-4.5,10.1-10.1,10.1c-5.6,0-10.1-4.6-10.1-10.1c0-5.6,4.5-10.1,10.1-10.1C84.2,46.7,88.7,51.3,88.7,56.8z};
  }
}

\newcommand\orcidicon[1]{\href{https://orcid.org/#1}{\mbox{\scalerel*{
\begin{tikzpicture}[yscale=-1,transform shape]
\pic{orcidlogo};
\end{tikzpicture}
}{|}}}}

\begin{document}

\preprint{APS/123-QED}

\title{Full-Continuum Ultrafast Quantum Gas Microscopy}

\author{Andrea Litvinov\,\orcidicon{0009-0001-5332-1188}}
\thanks{These two authors contributed equally.}
\affiliation{Institut f\"{u}r Quantenoptik und Quanteninformation, \"Osterreichische Akademie der \\ Wissenschaften, Technikerstr. 21A, 6020 Innsbruck, Austria }

\author{Clemens Ulm\,\orcidicon{0009-0009-7589-9536}}
\thanks{These two authors contributed equally.}
\affiliation{Institut f\"{u}r Quantenoptik und Quanteninformation, \"Osterreichische Akademie der \\ Wissenschaften, Technikerstr. 21A, 6020 Innsbruck, Austria }
\affiliation{Universität Innsbruck, Fakultät für Mathematik, Informatik und Physik,
Institut für Experimentalphysik, 6020 Innsbruck, Austria}

\author{Eva Casotti\,\orcidicon{0000-0002-8340-1445}}
\affiliation{Institut f\"{u}r Quantenoptik und Quanteninformation, \"Osterreichische Akademie der \\ Wissenschaften, Technikerstr. 21A, 6020 Innsbruck, Austria }
\affiliation{Universität Innsbruck, Fakultät für Mathematik, Informatik und Physik,
Institut für Experimentalphysik, 6020 Innsbruck, Austria}

\author{Gerard Solà Berga\,\orcidicon{0009-0006-5439-8247}}
\affiliation{Institut f\"{u}r Quantenoptik und Quanteninformation, \"Osterreichische Akademie der \\ Wissenschaften, Technikerstr. 21A, 6020 Innsbruck, Austria }
\affiliation{Universität Innsbruck, Fakultät für Mathematik, Informatik und Physik,
Institut für Experimentalphysik, 6020 Innsbruck, Austria}

\author{Manfred J. Mark\,\orcidicon{0000-0001-8157-4716}}
\affiliation{Institut f\"{u}r Quantenoptik und Quanteninformation, \"Osterreichische Akademie der \\ Wissenschaften, Technikerstr. 21A, 6020 Innsbruck, Austria }
\affiliation{Universität Innsbruck, Fakultät für Mathematik, Informatik und Physik,
Institut für Experimentalphysik, 6020 Innsbruck, Austria}

\author{Francesca Ferlaino\,\orcidicon{0000-0002-3020-6291}}
\thanks{Correspondence should be addressed to: \mbox{\url{francesca.ferlaino@uibk.ac.at}}}
\affiliation{Institut f\"{u}r Quantenoptik und Quanteninformation, \"Osterreichische Akademie der \\ Wissenschaften, Technikerstr. 21A, 6020 Innsbruck, Austria }
\affiliation{Universität Innsbruck, Fakultät für Mathematik, Informatik und Physik,
Institut für Experimentalphysik, 6020 Innsbruck, Austria}
\affiliation{Donostia International Physics Center (DIPC), 
Paseo Manuel de Lardizabal 4, 20018 Donostia-San Sebasti\'{a}n, Spain}

\date{\today}

\begin{abstract}

Quantum gas microscopy provides single-shot access to many-body configurations, yet has so far relied on pinning atoms to a discrete lattice during fluorescence readout. Here, we realize ultrafast atom-resolved fluorescence microscopy of an ultracold $^{162}$Dy gas directly in free space, with all external confinement removed during imaging. Exploiting the large mass and broad optical transition of dysprosium together with a high numerical aperture in-vacuum objective, we detect single atoms within $4\,\mu\mathrm{s}$ while maintaining an effective point-spread function with a full width at half maximum of $254.8(3)\,\unit{nm}$. We achieve a single-atom conditional classification fidelity of $99.72(3)\%$ for identified fluorescence candidates. We use the measured single-atom response to construct a statistical configuration-inference model that recovers atomic positions and multiplicities, coming from overlapping atoms, directly from single fluorescence snapshots. We further benchmark our reconstruction algorithm through end-to-end Monte Carlo simulations with known ground-truth atomic configurations, obtaining a reconstruction fidelity of $98.7\%$. Our approach, applicable to a wide variety of atomic species, extends quantum gas microscopy beyond predefined detection lattices, enabling atom-resolved measurements of counting statistics and higher-order spatial correlations directly on the intrinsic length scales of continuum many-body systems.
\end{abstract}

\maketitle

Quantum gas microscopy has transformed the study of many-body systems by providing direct, atom-resolved snapshots of their microscopic configurations rather than only global observables\,\cite{Bakr2009aqg,Sherson2010sar}. This capability has revolutionized the study of lattice Hamiltonians, enabling e.\,g.\,the observation of phenomena ranging from string order and antiferromagnetic correlations to nonlocal pairing\,\cite{Endres2011ooc,Mazurenko2017aca,Hartke2023doo}. Crucially, in these systems both the many-body dynamics and the microscopic readout occur in discrete space. At readout, the particle positions are pinned in deep lattices or tweezers, providing a spatial prior for reconstructing each experimental snapshot\,\cite{Gross2021qgm,Kaufman2021qsw}.

For quantum gases in continuous space, the situation is fundamentally different. The atom positions themselves constitute continuous degrees of freedom and encode the microscopic structure of the many-body state. Direct access to these coordinates provides stringent tests of interacting quantum field theories and thermodynamic relations both in and out of equilibrium\,\cite{Ott2024hlq, Esposito2009nff, Polkovnikov2011ndo, Eisert2015qmb, Chalopin2026cci}. Atom-resolved measurements of quantum gases evolving in the continuum have enabled the study of density fluctuations, short-range correlations, and counting statistics\,\cite{Dixmerias2026ftq, Xiang2025isi, Yao2025mpc, deJongh2025qgm, Daix2026osc, Dixmerias2026urm}. Their fluorescence readout, however, still relies on projecting the atomic distribution onto a pinning lattice, thereby reintroducing a discrete detection basis at the measurement stage. Such a lattice projection may e.\,g.\,distort short-range correlations as their length scales approach the lattice spacing\,\cite{Xiang2025isi, Dixmerias2026ftq}, while light-assisted collisions can lead to parity-projected occupancies rather than direct atom-number readout\,\cite{Xiang2025isi, Gross2021qgm, Yao2025mpc}. These effects have motivated the use of dilute gases, with mean interparticle separations of several lattice spacings\,\cite{Yao2025mpc, deJongh2025qgm, Dixmerias2026ftq}, or low lattice-site occupations\,\cite{Xiang2025isi}.

Full-continuum quantum gas microscopy requires direct atom-resolved imaging of a many-body system in continuous space. The absence of a spatial prior makes this fundamentally more demanding than lattice or tweezer microscopy. The central challenge is measurement back-action: each scattered photon contributes to the detection signal but also imparts a stochastic recoil kick, progressively washing out spatial information during the exposure.

Pioneering demonstrations of single-atom-resolved fluorescence imaging in continuous geometries, including free space, were performed with alkali atoms, addressing either an individual atomic emitter or a small number of well-separated emitters\,\cite{Buecker2009sps,Fuhrmanek2010ias,Bergschneider2018srs}. In $^{87}\mathrm{Rb}$, a single atom could be detected in time-of-flight fluorescence imaging. However, reconstructing its spatial response required many 
 experimental repetitions due to a single-shot detected-photon rate of about $0.02 \,\text{photons}/\unit{\mu s}$\,\cite{Fuhrmanek2010ias}. 
 A major step forward was subsequently achieved with $^{6}\mathrm{Li}$, where a few individual unpinned atoms were detected\,\cite{Bergschneider2018srs}. Owing to the large recoil of lithium, however, the fluorescence signal broadened strongly during the $20\,\unit{\mu s}$ exposure, yielding a large atomic point-spread function: discriminating two atoms with ${>}\,90\%$ probability required separations exceeding $30\,\unit{\mu m}$. For comparison, this corresponds to roughly 60 sites of a conventional optical lattice with $532\,\unit{nm}$ spacing, remaining far from the interparticle length scales relevant to quantum-gas microscopy.

The favorable combination of large atomic mass and broad optical transitions in lanthanides was subsequently identified as a route to ultrafast microscopy without continuous cooling or pinning in deep lattice, motivating proposals for magnetic atoms at submicrometer separations in both continuous and discrete geometries\,\cite{Picard2020dla}. The discrete implementation has since been realized experimentally in optical lattices\,\cite{Su2023dqs,Su2025fsa} and optical-tweezer arrays with dipolar atoms\,\cite{Gruen2024ota,Gruen2026lac} and alkaline-earth-like species\,\cite{Ma2023hfg,Scholl2023eci,MuziFalconi2025msh,Plassmann2026rsr}, reaching high-fidelity fluorescence detection. In these architectures, readout is further facilitated by the large emitter spacing: submicrometre lattices are expanded to several micrometres using an accordion technique\,\cite{Su2023dqs,Su2025fsa}, while tweezer arrays intrinsically operate at micrometre-scale separations\,\cite{Kaufman2021qsw}.

The continuum counterpart has remained unexplored, and this is the regime addressed here. Using an ultracold $^{162}\mathrm{Dy}$ gas, we demonstrate atom-resolved ultrafast fluorescence microscopy in which atoms at sub-micrometre separations are imaged directly in continuous space without a pinning lattice or any predefined spatial prior.
We reach this regime by combining the rapid photon scattering and reduced recoil of dysprosium with the high fluorescence-collection efficiency of our in-vacuum objective. At an illumination time of $4\,\unit{\mu s}$, we observe a single-emitter response that remains within $5\%$ of the diffraction limit ($\simeq 243\, \unit {nm}$) while retaining near-unity single-atom classification fidelity.

Importantly, our detection provides access to both single-emitter photon-count statistics and spatial information encoded in the measured point-spread function (PSF).
Combining this knowledge with a statistical geolocation-inspired multilateration approach allows us to go beyond the isolated-emitter regime: we first construct a spatial atom-number map that identifies candidate emitter positions and restricts the configuration space. We then evaluate the likelihood of the candidate atomic configurations directly against the measured fluorescence, including multiplets that contribute within a single diffraction-limited feature, to infer their most probable positions and multiplicities.
In this way, information that is inaccessible to conventional peak detection is recovered statistically from the joint spatial and photon-number information encoded in the fluorescence image, opening up a route to atom-number- and configuration-resolved microscopy of many-body systems directly in continuous space.

\begin{figure}[t!]
    \centering
    \includegraphics[width=\columnwidth]{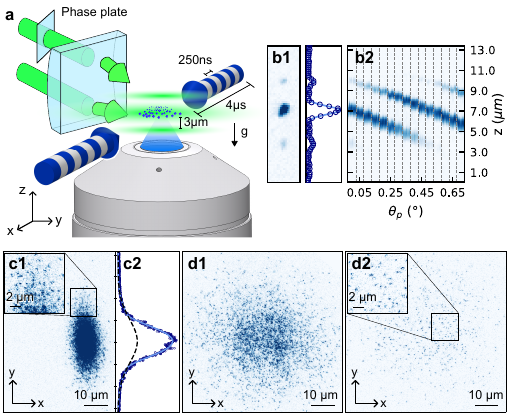}

    \caption{\textbf{In-vacuum quantum-gas microscope.}
        \textbf{a} Trapping and imaging protocol. Atoms are loaded in a single plane of a vertical accordion lattice with $3\,\unit{\mu m}$ spacing, whose phase is controlled by rotating a phase plate. The vertical lattice provides a band gap exceeding $h\times3\,\unit{kHz}$, corresponding to a harmonic-oscillator length $a_{\rm ho}\,{<}\,150\,\unit{nm}$, while the in-plane trap frequencies range between $2\pi\times 5-50\,\unit{Hz}$.  We switch off all trapping beams $1\,\unit{\mu s}$ before imaging. For imaging, the atoms are illuminated with resonant counter-propagating beams alternated in $250\,\unit{ns}$ pulses, and fluorescence is collected by the high-NA in-vacuum objective.
       \textbf{b} Single-layer loading and magnifier. \textbf{b1} Representative absorption image after quantum-gas magnification by a factor of 28.5, resolving individual lattice planes (left) and corresponding integrated profile (right). \textbf{b2} In-situ atomic position versus phase-plate rotation angle; each column contains 10 integrated profiles. \textbf{c}-\textbf{d} Representative fluorescence images with $4\,\unit{\mu s}$ illumination time for a BEC (c) and an ultracold gas (d). The BEC image (c1) is shown with its corresponding integrated profile and bimodal fit (c2) (solid line: total fit; dashed line: thermal component). For (d), we show examples for both a dense (d1) and sparse (d2) gas.}
    \label{fig:Fig1}
\end{figure}

\section{\label{sec:level2_new} Ultrafast microscopy in continuous space}

\noindent The first key ingredient enabling ultrafast microscopy in continuous space is provided by the favorable atomic properties of lanthanides\,\cite{Picard2020dla}. Their large mass suppresses recoil-induced motion, while broad optical transitions enable many photons to be scattered within a short exposure. For $^{162}\mathrm{Dy}$, we image on the broad $421\,\unit{nm}$ transition $(\Gamma/2\pi\,{=}\,32\,\mathrm{MHz})$, several times broader than typical alkali D lines and corresponding to a saturated scattering rate of $101$ photons/$\unit{\mu s}$. The short wavelength of this transition further favors high spatial resolution, with the characteristic resolution scale ${\propto}\,\lambda/\mathrm{NA}$, where NA is the numerical aperture.
The second key ingredient is efficient fluorescence collection. Our custom in-vacuum objective features an NA of $0.89$, corresponding to a collection angle of $128^\circ$, allowing the collection of about $30\%$ of the scattered photons\,\cite{Sohmen2023asi}; its main specifications are summarized in Appendix\,\ref{app:ImagingEffic}.

The experimental sequence is detailed in Appendix\,\ref{app:Expsequence}. In brief, $^{162}\mathrm{Dy}$ atoms are initially laser cooled and loaded into an optical transport trap in a first vacuum chamber, before being optically transferred into a custom quartz science cell housing the in-vacuum microscope objective. Figure\,\ref{fig:Fig1}a provides a schematic overview of our quantum-gas-microscope setup. There, we first perform evaporative cooling, bringing the gas close to or into quantum degeneracy, before adiabatically loading the atoms into a single plane of a vertical lattice, realizing a quasi-two-dimensional geometry. Single-layer loading, which we verify using horizontal absorption imaging combined with matter-wave magnification\,\cite{Asteria2021qgm,Murthy2014mvf} (Fig.\,\ref{fig:Fig1}b), is an important step to confine the gas within the depth of focus. However, the quasi-two-dimensional confinement is used only for sample preparation: immediately before fluorescence imaging, all external confinements are switched off within $1\,\mu\mathrm{s}$, and the atoms evolve in free space throughout the microscopic readout.

Fluorescence is induced by illuminating the atoms with two counter-propagating resonant beams, operating on the broad $421\,\unit{nm}$ transition. The counter-propagating configuration helps to counteract the momentum transfer due to radiation pressure, limiting atomic diffusion during imaging. Moreover, the beams are alternatively pulsed for $250\,\unit{ns}$ to avoid the creation of standing waves\,\cite{Su2023dqs,Anich2024cco,Su2025fsa}, while the total illumination time, $\tau$, is varied between $2$ and $8\,\unit{\mu s}$.
Photons scattered by the atoms are collected by the microscope objective and subsequently focused onto a camera.

The capability of our microscope is directly illustrated in Fig.\,\ref{fig:Fig1}c-d, which shows $4\,\unit{\mu s}$ single-shot fluorescence images of extended quantum gases resolved at the level of individual atoms. Figure\,\ref{fig:Fig1}c shows a Bose--Einstein condensate (BEC), for which the atom-resolved signal reproduces the smooth Thomas--Fermi envelope while simultaneously revealing microscopic density variations around it. These variations illustrate the direct access to local density fluctuations enabled by continuum microscopy, which will provide a natural observable for future studies of e.\,g.\,many-body correlations. Figure\,\ref{fig:Fig1}d shows near-degenerate samples at different densities, demonstrating that individual fluorescence features remain directly visible across a broad range of local densities and interparticle separations. 

\section{\label{sec:level3_new} Single-atom detection and spatial resolution in the continuum}

\noindent The images above establish atom-resolved fluorescence imaging of extended, many-body gases directly in continuous space on a microsecond timescale. As shown, the atomic emitters can occupy arbitrary positions in the imaging plane, so detection requires identifying an atomic emitter, that is, simultaneously localizing it and discriminating its fluorescence signal from background noise. This situation differs fundamentally from lattice quantum-gas microscopes, in which the detection problem is reduced to determining the occupancy of a fixed, discrete set of known trapping sites\,\cite{Gross2021qgm,Kaufman2021qsw}.

\begin{figure}[ht!]
\centering    
\includegraphics[width=1\columnwidth]{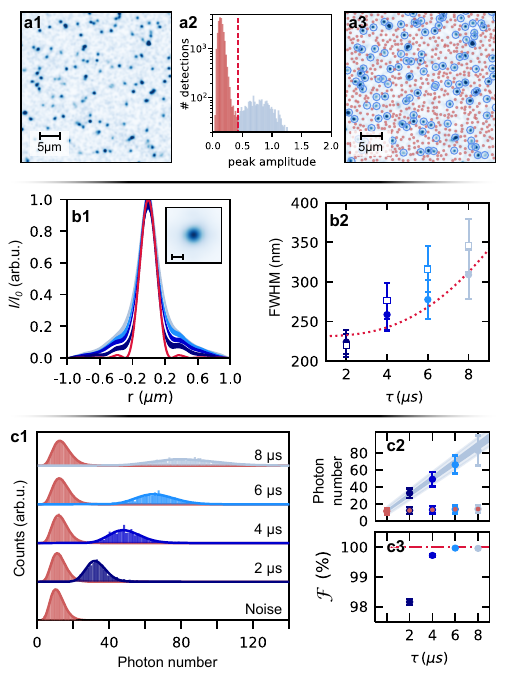}
    \caption{\textbf{Characterization of an ultrafast quantum gas microscope.}
    \textbf{a} Single-atom detection and classification: 
    \textbf{a1} Fluorescence image ($\tau\,{=}\, 4\,\unit{\mu s}$), after Gaussian filtering ($\sigma\,{=}\,245$\,\unit{nm}) for peak finding. The mean interatomic distance is $3.9\,\unit{\mu m}$. The projected pixel size is $122.7\,\unit{nm}$.
    \textbf{a2} Peak-amplitude distributions accounting for noise (pink) and atomic emitters (blue) compiled from over 27000 detected peaks. The dashed line marks the classification threshold.
    \textbf{a3} Detection classification overlaid on the image in a1: noise (pink dots) and atoms (blue circles, $2\,\unit{\mu m}$ diameter). 
    \textbf{b} Single-emitter PSF:
    \textbf{b1} Measured PSF profile along the direction perpendicular to the imaging beams for $\tau\,{=}\,2,\,4,\,6,\,8\,\unit{\mu s}$ from dark to light blue, respectively, together with the expected diffraction-limited Airy profile for our $\mathrm{NA}\,{=}\,0.89$ (solid red line).  Inset: supersampled PSF for $4\,\unit{\mu s}$; scale bar, $200\,\unit{nm}$. PSFs are averaged from 1400 isolated atoms, RMS errors are shown by the line thickness.
    \textbf{b2} FWHM of the effective PSF versus $\tau$. Filled and open symbols indicate directions perpendicular and parallel to the imaging beams, respectively. Error bars: propagated from b1. The red line is the expected transverse recoil-induced diffusion.
    \textbf{c} Photon-count statistics and detection fidelity:
    \textbf{c1} Photon-count histogram for different $\tau$ for noise (pink) and atoms (blue). Solid lines are gamma distribution fits. Histograms built from 1400 isolated atoms. The noise only histogram is built from images without atoms ($\tau = 4\,\unit{\mu s}$).
    \textbf{c2} Mean photon number for regions classified as atoms (blues) and noise (pink). Error bars: RMS spread. The light shaded band is the expected scaling of the mean and RMS of a Poissonian process. The dark shaded band is a linear fit with $\pm1\sigma$ uncertainty: 9.6(4) photons/$\unit{\mu s}$.
    \textbf{c3} $\mathcal{F}$ for isolated emitters, obtained from the overlap of the fitted histograms. Error bars: Monte Carlo bootstrapping.} 
    \label{fig:Fig2}
\end{figure}

We now quantitatively characterize the single-atom fluorescence response, following the sequence summarized in Fig.\,\ref{fig:Fig2}. We first identify candidate atomic emitters directly from the fluorescence images without imposing any positional prior (a). From sufficiently isolated emitters, we then reconstruct the effective single-atom PSF and characterize its dependence on the illumination time (b). Finally, we extract the corresponding photon-count statistics to quantify the single-atom classification fidelity and determine the optimal imaging time (c).

In particular, we begin by identifying isolated emitters. To this aim, we first Gaussian-filter the raw fluorescence image to smooth out pixel-scale fluctuations (Fig.\,\ref{fig:Fig2}a1), and then run a peak-finder algorithm to identify all local amplitude maxima, irrespective of their origin. As shown in Fig.\,\ref{fig:Fig2}a2, the resulting peak-amplitude distribution is clearly bimodal, separating a low-amplitude population of noise-induced maxima from a higher-amplitude population associated with atomic fluorescence. We define an amplitude threshold at the local minimum between the two populations (dashed line), which we use to discriminate the atomic emitters among all detected maxima. The resulting classification is shown in Fig.\,\ref{fig:Fig2}a3.

In the next step, we use this classification to select isolated emitters directly from the raw images, defined here as those with a nearest-neighbor separation of at least $2\,\mu\mathrm{m}$. We then characterize their effective PSF as a function of $\tau$. Here, ``effective'' denotes the spatial response measured during the finite fluorescence exposure, which includes both the optical PSF and the residual broadening from photon-recoil-induced atomic motion\,\cite{Su2025fsa}.

For each $\tau$, we extract a $2\times2\,\mu\mathrm{m}^2$ region of interest (ROI) around each isolated atomic emitter from the raw image. Exploiting that emitter positions are not necessarily centered on camera pixels, we resample the ROIs onto a grid ten times finer, recenter, and average to reconstruct a supersampled single-emitter PSF\,\cite{Bakr2009aqg}. We show representative PSF profiles for different $\tau$ along the direction perpendicular to the imaging beams in Fig.\,\ref{fig:Fig2}b1.
The measured profiles closely follow the ideal Airy response expected for $\mathrm{NA}\,{=}\,0.89$ (solid red line), with clearly resolved Airy rings, indicating that the single-emitter response remains close to the diffraction limit.
As $\tau$ increases, the effective PSF broadens, as shown in Fig.\,\ref{fig:Fig2}b2. The increase in the transverse full width at half maximum (FWHM) is well described by a diffusive $\tau^{3/2}$ scaling (dash-dotted line), consistent with recoil-induced momentum spread from stochastic spontaneous emission\,\cite{Bergschneider2018srs}. Along the imaging-beam direction, the PSF is slightly broader due to residual absorption recoil along $\pm\mathbf{k}$. The counter-propagating beams cancel the mean radiation pressure, but not its fluctuation, leading to a stochastic momentum spread\,\cite{Su2025fsa}.

Finally, we extract the photon-counting statistics of the local maxima, i.\,e.\,irrespective whether they originate from atomic fluorescence or noise; as shown in Fig.\,\ref{fig:Fig2}c1. In particular, we measure the photon number by integrating the raw image within a circular region of interest of diameter $2\,\unit{\mu m}$ centered on each local maximum. Moreover, to eliminate crosstalk between noise and atomic signal and between multiple atoms in close proximity, we restrict the analysis to peaks separated by more than the integration diameter from their nearest atom.

The resulting photon-count distributions reveal two well-separated populations associated with noise-induced maxima (pink) and single-emitter fluorescence (blue). As $\tau$ increases, the single-emitter distribution shifts toward larger photon counts and simultaneously broadens, while the noise distribution remains unchanged. To quantify this behavior, we fit the noise and single-emitter populations with gamma distributions; see Appendix\,\ref{app:PhotonScatteringStatistics}. The extracted mean photon number increases linearly with $\tau$, consistent with fluorescence collection at a constant detection rate of $9.6(4)\,\mathrm{photons}/\mu\mathrm{s}$ per atom; see Fig.\,\ref{fig:Fig2}c2. Under saturated excitation, dysprosium atoms scatter $101$ photons/$\mathrm{\mu s}$, corresponding to an end-to-end photon-detection efficiency of $9.5(4)\%$, among the highest reported in literature. 

The overlap between the fitted noise and single-emitter photon-count distributions quantifies the probability of correctly classifying an isolated candidate maximum once it has been identified. We define this conditional classification fidelity as $\mathcal{F}$. As shown in Fig.\,\ref{fig:Fig2}c3, $\mathcal{F}$ sharply increases with $\tau$, rapidly reaching values above $99.5\%$ and then saturating around unity.
Based on the above results, we conclude that $\tau\,{=}\,4\,\mu\mathrm{s}$ provides the optimal balance between signal-to-noise ratio and spatial resolution. At this operating point, we collect about 40 photons per atom with $\mathcal{F}\,{=}\,99.72(3)\%$ and a perpendicular FWHM of $254.8(3)\,\mathrm{nm}$. The latter is only about $5\%$ above the diffraction-limited value of ${\simeq}\,243\,\mathrm{nm}$ expected for $\lambda\,{=}\,421\,\mathrm{nm}$ and $\mathrm{NA} \,{=}\,0.89$.

It is useful to put these numbers in perspective. First, our measured FWHM is less than half of the $532\,\mathrm{nm}$ site spacing commonly used in optical-lattice quantum gas microscopes, placing the spatial response of our continuum microscope well below this characteristic length scale. Second, the closest previous demonstration of unpinned single-atom fluorescence imaging in free space, performed with $^{6}\mathrm{Li}$, detected approximately 25 photons in $20\,\mu\mathrm{s}$, corresponding to ${\sim}\,1.25$ detected photons/$\mu\mathrm{s}$, compared with a fitted rate of $9.6(4)$ photons/$\mu\mathrm{s}$ in our experiment\,\cite{Bergschneider2018srs}. More importantly, recoil-induced diffusion broadened the time-integrated single-atom signal in that experiment to a root-mean-square radius of $10.1(1.4)\,\mu\mathrm{m}$, and two atoms could be distinguished with ${>}\,90\%$ probability only for separations exceeding $32.4\,\mu\mathrm{m}$. In contrast, our $4\,\mu\mathrm{s}$ readout preserves an effective single-emitter FWHM of $254.8(3)\,\mathrm{nm}$, within $5\%$ of the diffraction limit. 

Here, the performance of ultrafast imaging in the continuum is demonstrated with dysprosium atoms, exploiting their heavy mass and the broad linewidth of the imaging transition. Yet, a wide variety of atomic species such as lanthanides and alkaline-earth like atoms share similar properties, and the principles used in this paper naturally extend to these atoms with comparable performances (see Appendix\,\ref{app:diffSpecies}).

\begin{figure}[t!]
    \centering
    \includegraphics[width=1\columnwidth]{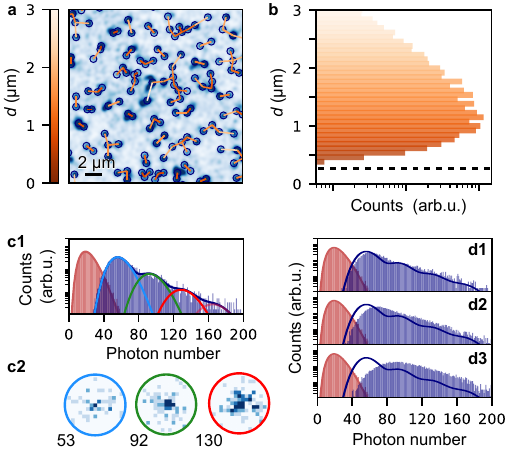}
    \caption{\textbf{Photon-counting statistics across inter-atomic distances.}
    \textbf{a} Blurred fluorescence image, $\tau\,{=}\, 4\,\unit{\mu s}$, with detected atom peaks connected by a nearest-neighbor (NN) graph (solid lines). The colormap encodes the NN distance $d_\text{NN}$. Here, the mean interparticle distance is $1.65\,\mu\mathrm{m}$.
    \textbf{b} Histogram of $d_\text{NN}$. The dashed line indicates the PSF FWHM as calibrated in Fig.\,\ref{fig:Fig2}b.
    \textbf{c1} Photon-count histogram for noise (pink) and atoms (blue), in the isolated emitter regime where $d\,{>}\, 2\,\unit{\mu m}$. Solid lines: fit of the histogram (see main text), accounting for multiple contributing atoms within one detected peak (blue: one atom, green: two atoms, red: three atoms, purple: four atoms).
    \textbf{c2} Sample multiplets of one (blue), two (green), and three (red) atoms, and corresponding integrated photon count below, corresponding to multiplet populations fitted in c1.
    \textbf{d} Photon-count histogram beyond the isolated emitter regime, where the range of NN distance of the peaks contributing to the histogram are $2\,\unit{\mu m}\,{>}\,d\,{>}\,1.5\,\unit{\mu m}$ (d1), $1.5\,\unit{\mu m}\,{>}\,d\,{>}\,1\,\unit{\mu m}$ (d2), $1\,\unit{\mu m}\,{>}\,d\,{>}\,0.5\,\unit{\mu m}$ (d3). The solid line is the fit result of c1, highlighting the histogram's shift towards higher photon count.}
    \label{fig:Fig3}
\end{figure}

\section{\label{sec:MultipleAtoms} Microscopy across interatomic length scales}

\noindent Next, we test our continuum microscope in a denser regime, where progressively shorter interatomic separations challenge number-resolved discrimination of emitters. 
In this regime, intensity maxima classified as atomic signal can come from a single or overlapping emitters, while the ones related to noise have already been filtered out following the procedure described in the previous section.
Figure\,\ref{fig:Fig3}a shows a representative fluorescence image together with the nearest-neighbor (NN) graph of the detected maxima, where each detection is connected to its nearest neighbor and the edge color encodes the corresponding NN distance, $d$.
We observe a continuous spectrum of $d$, as quantified in Fig.\,\ref{fig:Fig3}b, ranging from isolated maxima ($d\,{>}\, 2\,\unit{\mu m}$) down to peaks with sub-micrometer $d$, where dipolar interactions are expected to become relevant to many-body physics\,\cite{Baier2016ebh, Su2023dqs}.

For $d\,{>}\, 2\,\unit{\mu m}$, the photon-count histogram clearly shows that even a single isolated maximum can contain information from more than one emitter; see Fig.\,\ref{fig:Fig3}c1. In particular, once multiple emitters merge into a single fluorescence maximum, their increased photon count still carries information about the number of contributing atoms, as illustrated by the identified multiplets in Fig.\,\ref{fig:Fig3}c2.
We fit to the data a weighted sum of $n$-fold convolutions of the single-atom photon-count distribution with noise, accounting for $n\,{=}\, 1, 2, 3, 4$ atoms per detected peak. We observe good agreement with the fit, confirming that our fluorescence measurement conserves the atom-number information\,\cite{Picard2020dla, Su2025fsa, Gruen2024ota, MuziFalconi2025msh} rather than reducing the signal to a binary measurement of parity-projected occupancy.

We further investigate the photon-counting statistics while releasing the constraint on the large distance between the evaluated peaks. Figure\,\ref{fig:Fig3}d shows that when $d$ is reduced, the photon-count histograms are shifted toward higher photon numbers, as highlighted by the comparison with the fit of the isolated-peak histogram (solid line).

This is a clear signature that cross-talk between neighboring peaks occurs. In this regime, the total detected light at any given peak reflects contributions from multiple overlapping fluorescence spots simultaneously, obscuring the individual atomic contributions and and preventing the unambiguous extraction of single-atom positions from the integrated signal alone. This calls for the development of a new approach able to account for the joint spatial and photon-count statistics of multiple atoms simultaneously at arbitrary distances, as presented in the following section.

\begin{figure*}[ht!]
\includegraphics[width=1\columnwidth]{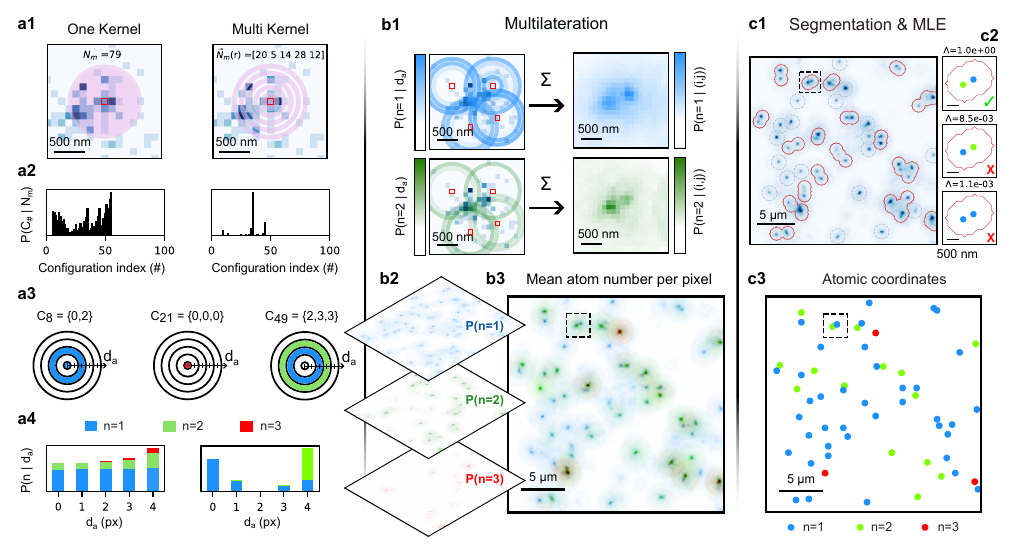}
    \caption{
        \textbf{Statistical deconvolution and reconstruction of the full atomic configuration.}
\textbf{a} Local radial inference. \textbf{a1} The photon number $N_m$ is measured in radially symmetric kernels $K$ centered on a reference pixel, using either a single or multiple concentric $K$ of different radii $r$. \textbf{a2} Conditional probability $P(C_\#|N_m)$ of the candidate radial configurations $C_\#$, obtained from a single kernel (left) and from multiple kernels (right). \textbf{a3} Examples of radial configurations, specified by the atom distances $d$ from the reference pixel. \textbf{a4} Corresponding atom-number probabilities $P(n|d)$ as a function of $d$. \textbf{b} Multilateration. \textbf{b1} Radial atom-number probabilities inferred from different reference pixels are projected onto the image plane and combined to recover the two-dimensional spatial information. \textbf{b2} Resulting number-resolved spatial maps for atomic multiplicities $n\,{=}\,1,2,3$. \textbf{b3} Mean atom number per pixel, $\langle n\rangle(i,j)$, obtained from the number-resolved maps. \textbf{c} Final configuration reconstruction via Maximum Likelihood Estimator. \textbf{c1} Local maxima of $\langle n\rangle(i,j)$ define candidate atomic positions and constrain their multiplicities. \textbf{c2} Nearby candidates are clustered and the corresponding configurations $C_\#$ are evaluated against the raw fluorescence data through the likelihood $\Lambda(C_\#)$. \textbf{c3} Reconstructed atomic coordinates and multiplicities.  }
    \label{fig:Fig4}
\end{figure*}

\section{\label{sec:SpectralDeconvolution} Reconstruction of the atomic configuration}

\noindent Reconstructing the underlying many-body atomic configuration from a fluorescence image is the main purpose of quantum gas microscopy. When emitters are well separated, or when operating in a high-photon-number regime with atoms constrained to a predefined set of sites, conventional reconstruction and deconvolution methods provide a powerful route to determining atomic occupations\,\cite{Impertro2023aud, LaRooij2023acs}. Our continuum microscope operates in a fundamentally different regime: a single fluorescence snapshot contains an unknown number of simultaneously emitting atoms, whose positions and mutual separations are arbitrary and whose spatial responses can strongly overlap, while only a few tens of photons are detected from each emitter. Similar inverse problems arise in crowded-field stellar photometry and single-molecule localization microscopy, where overlapping point sources at unconstrained positions are typically disentangled using multiple frames or repeated measurements\,\cite{Stetson1994tco,Lelek2021sml}. Here, however, the objective is to reconstruct the complete many-body atomic configuration of a single experimental realization.

A distinctive advantage of the atomic system is that every emitter is physically identical and its response can be independently characterized. The measurements of the preceding sections provide both the spatial response of a single atom through the PSF and its photon-count statistics. We use this information to formulate the reconstruction as a statistical configuration-inference problem, rather than an intensity-deblurring problem. The central question is: among all possible sets of atomic positions, how likely is each configuration to account for the observed photon pattern? For any configuration of atomic positions, we analytically construct the probability distribution of the photon count. This forward statistical model for the fluorescence signal can then be inverted to compare the likelihoods of competing atomic configurations. The complete derivation is given in Appendix\,\ref{app:deconvolutionmodel}.

An additional challenge is the size of the configuration space. Whereas a lattice restricts the possible emitter positions to a finite set of sites, the number of configurations in continuous space rapidly becomes intractable. Our key simplification is to exploit the radial symmetry of the single-atom response together with annular measurement kernels: relative to any chosen kernel center, a local atomic configuration can be represented only by the radial distances of its constituent atoms, reducing the local reconstruction to a finite set of one-dimensional radial configurations. Repeating this inference from many centers across the image and projecting the resulting radial information onto the two-dimensional plane performs a multilateration, yielding spatial maps for the probability of different atom numbers. Although this result already maps the measured photon intensity to the underlying atomic density, we extend this technique to extract atomic positions. The deconvolved atomic density is used to restrict the otherwise enormous configuration space to a small set of candidate positions, whose configurations are finally evaluated against the raw fluorescence data to determine the most likely positions of all detected atoms. 

Figure\,\ref{fig:Fig4} presents this statistical reconstruction in three successive stages: local radial inference, two-dimensional multilateration, and final configuration selection. We briefly describe below the main steps, while a full discussion is given in Appendix\,\ref{app:algorithmImplementation}.

In the first step, we choose a reference pixel in the fluorescence image and determine which local atomic configurations are compatible with the photon number measured around it (see Fig.\,\ref{fig:Fig4}a1--a4). We quantify the signal by integrating the image over regions of pixels (kernels $K$). For a measured photon number $N_m$ in a given kernel, our model allows us to analytically evaluate the probability of every candidate atomic configuration $C_\#$, $P(\vec{N}_m|C_\#,K)$. Using a Bayesian inversion with a flat prior, we directly obtain the conditional probability of each configuration given the reference pixel. We choose $K$ to be radially symmetric, using concentric annular regions of different radii $r$. The atomic configurations then depend only on the radial distance $d$ from the reference pixel, reducing the local configuration problem to one dimension (a3). Finally, marginalizing over configurations yields the probability of finding a given number of atoms at each radial distance from that pixel (a4). While a single annulus generally leaves many configurations compatible with the measurement, Fig.\,\ref{fig:Fig4}a shows that combining the information from several radii strongly constrains the allowed configurations (a2, a4).

In the second step, we recover the two-dimensional spatial information lost in the radial reduction (b1--b3). The radial inference around a single kernel center constrains only $d$ of an emitter, but not its angular position, producing ring-like probability distributions. Repeating the same inference from multiple kernel centers across the image, and projecting the radial information onto the image plane, makes these constraints intersect, performing a multilateration (b1). This produces not only the expectation value of the atom number per pixel (b3), but also the averaged marginal probability per pixel for each multiplicity (b2). This methodology naturally suppresses both background and photon shot noise, and gives us precise knowledge of atomic positions via local probability maxima.

In the final step, we convert these spatial maps into a specific atomic configuration (c1--c3). Local maxima of $\langle n\rangle(i,j)$ define the candidate atomic positions, while their values constrain the corresponding multiplicities, thereby reducing the original configuration space to a tractable set (c1). The image is segmented, whereby nearby candidates are clustered, and the resulting configurations $C_\#$ are evaluated against the raw fluorescence data (c2). Importantly, the multilateration map is used only to identify and restrict the candidate configuration space, while the final likelihood $\Lambda(C_\#)$ is obtained directly from the measured fluorescence with a Maximum Likelihood Estimation (MLE). The configuration maximizing $\Lambda(C_\#)$ provides the reconstructed atomic coordinates and their multiplicities (c3).

To benchmark the complete reconstruction against a known ground truth, we apply the full algorithm to Monte Carlo fluorescence images generated with the same atomic density and experimentally characterized imaging response as our measurements (see Appendix\,\ref{app:montecarlo}). Because the underlying atomic positions are known a priori, the reconstructed configuration can be compared directly with the injected one. We obtain a total reconstruction fidelity of $98.7\%$, demonstrating high-fidelity recovery of many-body atomic configurations under experimentally relevant conditions. Correctly identified atoms are localized to a RMS error of just $85\,\mathrm{nm}$.

A key strength of this approach is that the model is purely analytical and derived from first principles, without any training data. This opens straightforward and well-defined avenues for improvement, both in the reconstruction fidelity and algorithm efficiency, by incorporating more advanced techniques such as iterative deconvolution\,\cite{LaRooij2023acs}. \comment{}

\section{\label{sec:level6}Conclusion}

\noindent Our continuum quantum gas microscope establishes new measurement capabilities for quantum gases with broad applicability. First, the approach is not restricted to bosonic dysprosium but can be extended to other lanthanides, alkaline-earth-like and species with similar atomic properties, including their fermionic isotopes. Second, it provides direct access to the microscopic configurations of bulk many-body systems at interatomic separations well below conventional lattice spacings, extending quantum gas microscopy beyond lattice-based readout. This may open radically new possibilities for probing correlated quantum gases through few- and many-body correlations, full counting statistics, and microscopic structure, ranging from short-range pairing and the real-space correlations of quantum Hall fluids to emergent spatial order, topological defects, BKT physics, and turbulent dynamics\,\cite{Bloch2008mbp,Ketterle2008mpa,Hadzibabic2011,Dalibard2011agp,Madeira2020qtb}. More generally, continuum atom-resolved microscopy offers a new experimental window onto statistical physics at the microscopic level, providing access to higher-order cumulants, connected correlations, non-Gaussian fluctuations, and local thermodynamic observables\,\cite{Esposito2009nff,Polkovnikov2011ndo,Eisert2015qmb, Chalopin2026cci}.

\section*{Acknowledgements}
We would like to acknowledge Maximilian Sohmen for his work on designing the microscope and Matthew Norcia, Lauritz Klaus, and Claudia Politi for early experimental work, in particular on adding the glass vacuum chamber to the experiment. We thank Emil Kirilov for fruitful discussions. This work was supported by the European Research Council through the Advanced Grant DyMETEr (\href{https://doi.org/10.3030/101054500}{No.\,101054500}), the NextGeneration EU grant AQuSIM through the Austrian Research Promotion Agency (FFG) (No.\,FO999896041), and the Austrian Science Fund (FWF) Cluster of Excellence QuantA (\href{https://doi.org/10.55776/COE1}{10.55776/COE1}).
This research was also funded in part by the Austrian Science Fund (FWF) Grant (\href{https://doi.org/10.55776/F101200}{10.55776/F101200}).

\section*{Data availability}
Data to support these findings are available from the authors upon reasonable request.

\clearpage

\renewcommand{\figurename}{Fig.}
\renewcommand\thefigure{A\arabic{figure}} 
\renewcommand\thetable{A\arabic{table}} 
\setcounter{figure}{0}
\appendix

\section{Experimental Sequence \label{app:Expsequence}}

\noindent Our experimental apparatus comprises two vacuum chambers: a stainless-steel chamber dedicated to initial cooling and trapping, and a second, custom-made quartz glass chamber hosting a high-NA in-vacuum microscope objective\,\cite{Sohmen2023asi}, hereafter referred to as the science chamber. In the first chamber, approximately $10^7$ spin-polarized $^{162}$Dy atoms are captured and cooled to about $12\,\unit{\mu K}$ in a narrow-line magneto-optical trap (MOT)\,\cite{Ilzhoefer2018tsf}. They are subsequently loaded into a large ``reservoir'' beam that funnels atoms from the MOT into the tighter transport beam\,\cite{Wenzel2015adq}, which is simultaneously ramped up linearly in $80\,\unit{ms}$.
After about $300\,\unit{ms}$, the magnetic field direction is rotated in $20\,\unit{ms}$ to be aligned with the transport axis, while the reservoir trap is ramped off in $250\,\unit{ms}$. By translating the focus of the transport beam, the atoms are transported over $41\,\unit{cm}$ into the science chamber in $2\,\unit{s}$.

Once in the science chamber, the atoms are transferred to a large elliptical dipole trap (ODT\,1, $\lambda\,{=}\,1064\,\unit{nm}$, $w_x \,{\approx}\, 100\,\unit{\mu m}$, $w_z \,{\approx}\, 25\,\unit{\mu m}$) counter-propagating with respect to the transport beam. The power of this trap is ramped up linearly in $250\,\unit{ms}$. After a hold time of $100\,\unit{ms}$, the transport beam is ramped down exponentially in $150\,\unit{ms}$. The magnetic field is then instantaneously restored to the vertical direction, after which evaporation begins. In a first evaporation step, we exponentially ramp down the power of ODT\,1 for $250\,\unit{ms}$. In a second step, while continuing the evaporation with ODT\,1, we also ramp up a second ODT (ODT\,2, $\lambda\,{=}\,1064\,\unit{nm}$, $w_y\,{\approx}\, 30\,\unit{\mu m}$, $w_z \,{\approx}\, 30\,\unit{\mu m}$) in $500\,\unit{ms}$. The ODT\,2 beam propagates orthogonal to the transport direction, resulting in a crossed ODT (cODT). Finally, the power of both ODT beams is further reduced exponentially. 

The depth of the final trap and the speed of evaporation vary the final temperature and atom number (see Fig.\,\ref{fig:Fig1}c-d). A slow evaporation of about ten seconds produces degenerate gases with up to \num{20000} atoms with a BEC fraction larger than \num{90}\%. Fast evaporations of around two seconds allow us to produce sparse samples with temperatures ranging from $200\,\unit{nK}$ to $500\,\unit{nK}$ where the interparticle distance is large enough to perform the characterization of the microscope on single atoms, as presented in this manuscript. After evaporation, we adiabatically load the vertical lattice in $500\,\unit{ms}$, ramping up the power following an S-shaped curve. Subsequently, to further reduce the in-plane confinement, the cODT can be ramped down exponentially in an additional $500\,\unit{ms}$, depending on the target density.

Immediately after turning off all the traps within $1\,\unit{\mu s}$, we image the gas in free space using fluorescence imaging through our quantum gas microscope: counter-propagating beams pulsed alternately every $250\,\unit{ns}$ avoid the formation of standing waves and suppress atomic diffusion, keeping the PSF close to the diffraction limit; see Fig.\,\ref{fig:Fig2}b. Recoil-induced diffusion along the vertical direction contributes only at the percent level to the PSF broadening for pulse durations $\tau\,{<}\,10\,\unit{\mu s}$, while free-space expansion of the wavefunction is negligible. Alternatively, we image the atoms after time-of-flight (TOF) via horizontal absorption imaging. Before TOF, we can apply a quantum gas magnifier protocol to resolve the lattice sites, see Fig.\,\ref{fig:Fig1}b.

\section{Optical Transport \label{app:OpticalTransport}}

\begin{figure}[]
\centering    
\includegraphics[width=1\columnwidth]{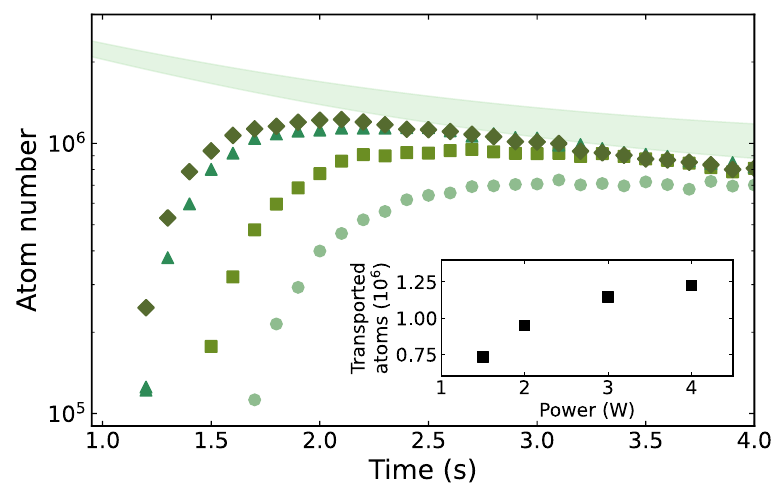}
\vspace{-0.0cm}
    \caption{\textbf{Transport efficiency and trap lifetime.} Atom number measured in the science chamber via absorption imaging as a function of transport duration for various laser powers: $1.5\,\unit{W}$ (circles), $2\,\unit{W}$ (squares), $3\,\unit{W}$ (triangles), and $4\,\unit{W}$ (diamonds). For long transport durations, the atom loss is governed by the tight-trap lifetime measured in the MOT chamber. The shaded region represents an exponential decay fit to the $3\,\unit{W}$ lifetime measurement ($\tau\,{=}\, 1.51(7)\,\unit{s}$), with the band envelope spanning the fits across all laser powers.}
    \label{fig:SupMatTP}
\end{figure}

\noindent To transport the atomic cloud $41\,\unit{cm}$ from the MOT to the science chamber, we employ a single-beam dipole trap ($\lambda\,{=}\,532\,\unit{nm}$) with a variable focal position. A hollow corner-cube retroreflector (Newport UBBR2.5-1S) on an air-bearing translation stage (Aerotech ABL1500-500), sitting on a granite block, is placed in the beam path between two $f\,{=}\,500\,\unit{mm}$ lenses forming a 1:1 telescope. 
The corner cube geometry guarantees that the reflected beam is perfectly parallel to the incoming beam, regardless of the orientation of the retroreflector, decoupling the beam from translation stage vibrations. Additionally, the air-bearing stage allows for very smooth and almost vibration-free motion. An additional achromatic doublet lens placed at $2f$ from the last telescope lens focuses the beam to a waist of $19.7(3)\,\unit{\mu m}$ in the vacuum chamber. If the retroreflector is positioned such that the beam path between the telescope lenses is $2f$, the light arrives collimated on the focusing lens, resulting in a focus at a distance $f$ from the lens. If the corner cube is displaced by a distance $d$, the beam path will be modified by $2d$. Consequently, the beam diameter and degree of collimation incident to the focusing lens will differ, resulting in a displacement of the focus position of $2d$.
Importantly, the beam waist remains constant throughout the transport, ensuring the atoms experience the same trap frequencies at all positions\,\cite{Leonard2014oto}. The corner cube is translated over a distance $d$ in time $t$ by linearly ramping the acceleration to $8d/t^2$ over $t/4$ before linearly decreasing it to $0$ over another $t/4$. This acceleration profile smoothly displaces the translation stage by $d/2$, reaching a maximum velocity of $2d/t$. Subsequently, the translation stage is decelerated in the same way.

The parameter governing the maximum speed of transport is the axial trap frequency of the transport beam, depending on the Rayleigh length, atomic mass, and total optical power. With a measured Rayleigh length of $1.5(3)\,\unit{mm}$ and a total power of $3\,\unit{W}$, we calculate a frequency of $20\,\unit{Hz}$. Higher axial trap frequencies allow for larger acceleration of the beam focus without spilling the atoms. We directly observe this effect when we investigate the number of atoms successfully transported as a function of the transport beam power and transport time; see Fig.\,\ref{fig:SupMatTP}.
The shaded light green area represents the lifetime of the atoms in the transport beam without displacing the focus, which is essentially independent of the chosen trap power within our experimental uncertainties. It is apparent that as the beam power is increased, the minimum time for successful transport is reduced. As the atoms spend less time confined in the tight dipole trap, this reduced transport time also reduces three-body collisions, yielding fewer losses during transport. Transport using $4\,\unit{W}$ of power yields more than $10^6$ atoms in the science chamber.

\section{Vertical Lattice\label{app:vertLattice}}

\noindent To fully exploit the resolving power of our imaging system, we need to confine the atomic wavefunctions such that they are smaller than the depth of focus, in our case $250\,\unit{nm}$. For this, we employ a lattice generated by the interference of two $532\,\unit{nm}$ beams incident at opposite angles $\theta$ with respect to the horizontal plane. We can tune the power of the beams to achieve a bandgap of up to $3.65\,\unit{kHz}$, corresponding to a depth of approximately $3000\,E_R$,  and to a harmonic oscillator length inside the central potential well of about $130\,\unit{nm}$. We add an anti-reflection-coated window (thickness $1\,\unit{mm}$), mounted on a piezo rotation stage, into the path of one of the beams. This allows us to achieve sub-wavelength adjustments of the relative path length, adjusting the phase of the interference pattern and therefore the vertical position of the intensity maxima, as demonstrated in the main text in Fig.\,\ref{fig:Fig1}b.

The vertical lattice is realized as follows. We use a 2" 50/50 beam splitter cube placed such that the coating is perpendicular to the vertical axis. A single horizontal incoming beam, such that the angle of incidence is $\pm\,45\degree$ to the entrance surface of the cube, is split into two horizontal beams, copropagating, with a vertical separation entirely determined by the height of the incident beam. The two outgoing beams are redirected to the gas with a single $f\,{=}\,70\,\unit{mm}$ achromatic doublet cylindrical lens, realizing the vertical lattice at their crossing. This design maximizes passive relative path-length stability, as the total path length between the beam splitter and the atoms is kept short.

While $\theta$ could be tunable to some degree depending on the experimental setup, we use a fixed value of about $\theta\,{=}\,5\,\unit{\degree}$, resulting in a lattice spacing of $a\,{=}\, \frac{\lambda}{2\sin(\theta)}\,{\approx}\, 3\,\unit{\mu m}$. We have chosen this specific value to allow the gas to be loaded into a single node of the optical lattice without the need to remove atoms in unwanted layers. We characterize this by directly measuring the population in the different lattice planes. As the resolving power of our horizontal absorption imaging system is too low to reliably distinguish lattice layers, we use a quantum gas magnifier scheme\,\cite{Murthy2014mvf,Asteria2021qgm}. Here, we employ the optical dipole trap previously used for evaporation (ODT\,1), allowing us to achieve a deep harmonic trap along the vertical axis with an expected frequency of $\omega\,{=}\,2\pi\cdot418\,\unit{Hz}$. In short, we switch off the lattice and let the atoms evolve in the dipole trap for a quarter period $T/4$ of the harmonic trap before a time-of-flight expansion of duration $T_\mathrm{TOF}$. The magnification of the system is then given by $M\,{\approx}\, \omega T_\mathrm{TOF}$ as discussed in Ref.\,\cite{Asteria2021qgm}. For our experimental parameters ($T/4\,{=}\,550\,\unit{\mu s}$, $T_\mathrm{TOF}\,{=}\,10\,\unit{ms}$), the resulting magnification is $28.6\times$, verified from the ratio of the observed layer separation to the known in-situ lattice spacing.

\begin{figure}[t]
\centering    
\includegraphics[width=1\columnwidth]{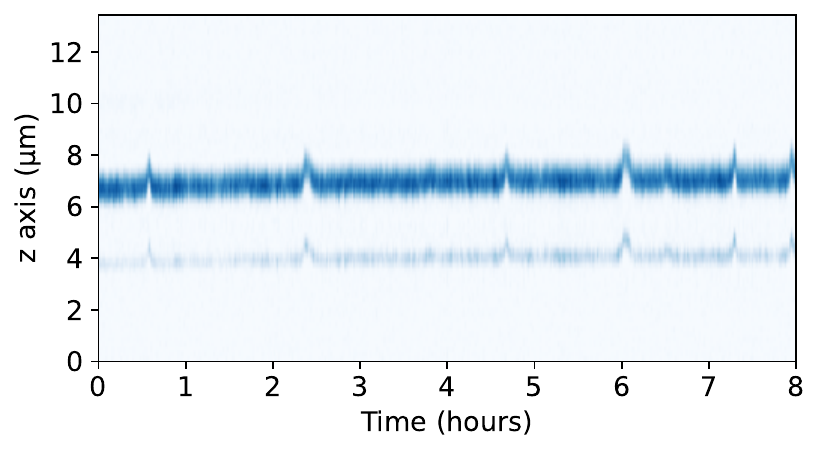}
\vspace{-0.0cm}
    \caption{\textbf{Vertical lattice stability measurement.}  Vertical lattice positional stability over 8 hours measured from absorption images after a matter-wave magnifier scheme and $10\,\unit{ms}$ TOF.}
    \label{fig:SingleLayerStability}
\end{figure}

This technique allows us to resolve the population in each of the lattice sites and maximize the single-layer loading as shown in Fig.\,\ref{fig:Fig1}b of the main text. This further suppresses all the noise originating from the out-of-focus atoms fluorescing in neighboring layers. The passive long-term single-layer loading stability over multiple hours is demonstrated in Fig.\,\ref{fig:SingleLayerStability}. Over an 8-hour period, the site populations remain remarkably constant with the exception of occasional fluctuations, which are easily filtered via post-selection. We speculate that those fluctuations are resulting from residual mode-jumps of the free-running laser source.

\section{Fluorescence imaging system and collection efficiency \label{app:ImagingEffic}}

\noindent We designed our fluorescence imaging system such that it is capable of detecting single atoms with a very low number of scattered photons per atom. The most relevant technical values of our high-NA objective are displayed in Table\,\ref{Tab:TechnicalDetails}.

\begin{table}[ht]
\centering
\begin{tabular}{|l | l|}
\hline
     Effective Focal Length & $20.0\,\mathrm{mm}$ \\
     Depth of Focus & $250\,\mathrm{nm}$ \\
     Numerical Aperture & $0.89$ \\
     Collection Efficiency & $27.0\%$ \\
     \hline
\end{tabular}
\caption{Specifications of our in-vacuum objective. The effective focal length (EFL) and the depth of focus (DOF) are extracted from simulations in ZEMAX OpticStudio, the NA was measured out-of-vacuum using a SNOM fiber\,\cite{Sohmen2023asi}. The collection efficiency takes into account the emission pattern of the scattered $\sigma^{+/-}$ polarized light, and a $\oslash\,1.5\,\unit{mm}$ mirror glued to the top of the objective.  Further technical details of our in-vacuum microscope design can be found in Ref.\,\cite{Sohmen2023asi}.}
\label{Tab:TechnicalDetails}
\end{table}

The light emitted by the atoms and collimated by the in-vacuum objective is transmitted through an anti-reflection-coated vacuum viewport and is subsequently directed onto a $750\,\unit{mm}$ achromatic doublet tube lens using a 3" dielectric mirror. The resultant magnification of $37.5\times$ is chosen such that the projected pixel size in the atomic plane ($122.7\,\unit{nm}$) is smaller than the optical resolution of the microscope, without spreading the PSF over too many pixels. Modern qCMOS cameras are not only single-photon sensitive but are also capable of resolving the quantitative number of photons incident on each pixel. Therefore, spreading the signal across pixels on the camera beyond the resolution limit reduces the signal-to-noise ratio, unlike sensors unable to count the photon number.

After passing through the f\,$750\,\unit{mm}$ tube lens, the light is directed towards a qCMOS camera. As our objective is placed in vacuum, we are unable to adjust the position of the object plane by scanning the objective height. Instead, we mount the tube lens on a motorized translation stage (Thorlabs MTS50/M-Z8), allowing us to fine-tune the image plane conjugation. Attached to the camera, we place two spectral filters (Thorlabs FESH0450 \& Semrock FF01-940-SP). Both the designed imaging wavelengths ($401\,\&\,421\,\unit{nm}$ for erbium and dysprosium, respectively) pass through the filters with minimal attenuation (${>}\,92\%$ transmission), while all other laser wavelengths used in the experiment are attenuated by at least \num{5} orders of magnitude.

\begin{figure}[t!]
\centering    
\includegraphics[width=1\columnwidth]{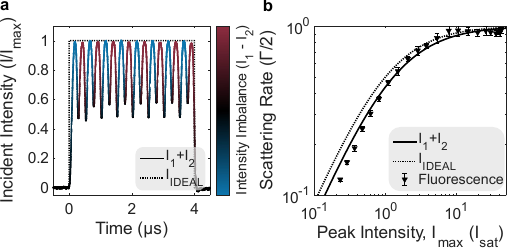}
\vspace{-0.0cm}
    \caption{\textbf{Imaging beams duty cycle and intensity.} \textbf{a} Imaging pulse train. The amplitude of the signal represents the total intensity incident on the atoms ($I_1+I_2$), while the color represents the intensity imbalance between the two pulsed beams ($I_1-I_2$). The dashed line represents an ideal square pulse shape. \textbf{b} The scattering rate as a function of peak intensity for the two pulse shapes illustrated in \textbf{a}. A measurement of atomic fluorescence as a function of varying peak intensity is fit to the curve, estimating our maximum intensity to be ${\approx}\,30\,I_{sat}$.}
    \label{fig:SupMatImaging}
\end{figure}

Right before imaging, all external confinements, i.\,e.\,the vertical lattice and, if still on, the cODT, are simultaneously turned off within $1\,\unit{\mu s}$. Then, counterpropagating imaging beams, pulsed alternately, saturate the $421\,\unit{nm}$ imaging transition. The measured intensity of both imaging beams, recorded on the monitoring photodiodes, is shown in Fig.\,\ref{fig:SupMatImaging}a alongside the ideal square pulse scenario. To ensure we drive the imaging transition above saturation, we scan the intensity of the imaging beams and observe the number of photons detected by the camera. Since the imaging light is pulsed rather than continuously driving the transition, we must account for the time during which the atom is not illuminated at maximum intensity. In Fig.\,\ref{fig:SupMatImaging}b, we plot the normalized scattering rate at different maximum beam intensities. The atomic-fluorescence measurement agrees well with the predicted scaling of the scattering rate for the pulsed beams. At our extracted maximum intensity, $30\,I_{sat}$, we estimate an additional $5\%$ loss due to the pulsed illumination profile.

\section{Statistics of camera noise and single-atom photon counts \label{app:PhotonScatteringStatistics}}

\noindent We characterize the single-pixel noise probability density function (PDF) by acquiring images with the camera sensor fully blocked. This noise PDF is intrinsic to the camera and captures contributions from readout noise, electronic noise, and related sources. When integrating the signal over a region of interest (ROI) centered on each detected peak, the resulting noise PDF is the $n$-fold convolution of the single-pixel noise PDF. For a circular ROI of $2\,\unit{\mu m}$ diameter, this corresponds to a 193-fold convolution, which yields a skewed distribution well described by a gamma function.

The signal from a single atom is governed by the number of photons it scatters during the imaging pulses, which follows a Poisson distribution. The measured single-atom PDF, i.\,e.\,the distribution of ROI sums around atomic positions, is therefore the convolution of this Poissonian photon distribution with the noise PDF.

For the statistical deconvolution, we employ the bare atomic photon distribution and consider the camera-noise contribution separately. 

\section{\label{app:deconvolutionmodel} Statistical deconvolution model}

\noindent The probability mass function (PMF) $P$ to detect $N$ photons on a pixel $(i,j)$ results from the contributions of all the atoms at emission range, i.\,e.\,with a non-zero overlap with their individual PSFs. Formally, the result is the convolution product of the individual and independent contributions:
\begin{equation}
    P(N) = P(N|\mathrm{noise}) \ast P(N|a_1) \ast ... \ast P(N|a_n),
\end{equation}
where the noise contribution is accounted for, and the $a_n$ are the contributing atoms indices. In the reciprocal space, this becomes a simple and tractable product:
\begin{equation}
    \phi(\omega) = \phi(\omega|\mathrm{noise}) \times \prod_{a} \phi(\omega|a),
\end{equation}
where $\phi$ are the characteristic functions of the PMFs. The statistical description of the measured photon number per pixel can then be entirely determined by the single-emitter contributions $\phi(\omega|a)$, and we will now show that given the single-atom PSF and photon PMF, we can derive a simple analytical expression. \\

The underlying physical processes of the single-emitter statistics can be broken down as follows. Each emitted photon has a spatial detection probability entirely described by the PSF centered on the atom at position $d$, and a region of interest $K$:
\begin{equation}
    p_K(d) = \int_{K}\mathrm{PSF}(d-r')dr'.
    \label{eq:PKD}
\end{equation}
Note that $K$ can be a single pixel, or a group of pixels, and for simplicity will be called kernel in the following. Each atom emits a random number of photons $N$, and the statistical description of photons detected in $K$ is rigorously a Bernoulli process. This Bernoulli process, $\mathcal{B}$, is a sequence of $N$ Bernoulli trials, where each detection event, $X\,{=}\,1$, happens with probability $p_K(d)$:
\begin{equation}
    \mathcal{B} = \sum_{i=1}^N X_i.
\end{equation}
It is important to note here that we assume that every photon is emitted independently, making $\mathcal{B}$ a compound distribution of independent $X_i$'s. To evaluate the characteristic function of a single emitter, $\phi_\mathcal{B}(\omega)$, with $N$ itself being a random variable, we use the law of total expectation by conditioning on $N$:
\begin{align}
    &\phi_\mathcal{B}(\omega \,|\, a) = E_N[\phi_X(\omega)^N \,|\, N]\\
    &\Rightarrow \phi_\mathcal{B}(\omega \,|\, a) = M_N[\log\phi_X(\omega)],
\end{align}
where $\phi_X(\omega)\,{=}\, E[e^{i\omega X}]$ and $M_N$ is the moment-generating function of $P(N)$. In particular, for the Poisson distribution with expectation value $\lambda$, this results in the analytic expression:
\begin{equation}
    \phi_\mathcal{B}(\omega \,|\, a) = \exp[\lambda (\phi_X(\omega)-1)].
\end{equation}
This formula entirely determines the contribution of a single emitter to the probability of detecting $N$ photons in the kernel $K$. Finally, for a given set of atoms, the PMF of photons in kernel $K$ is:
\begin{equation}
    P(N\,|\,K) = \mathcal{F}^{-1}[\phi(\omega\,|\,\mathrm{noise}) \times \prod_{a} \phi_\mathcal{B}(\omega \,|\, a)].
    \label{eq:PNK}
\end{equation}

\section{\label{app:algorithmImplementation} Algorithm implementation}

\noindent The implementation of Eq.\,\ref{eq:PNK} to detect the atomic positions requires defining a set of atomic configurations $C_\#$, for each given kernel $K$ at which the measured photon number $N_m$ is evaluated. In this work, we chose a set of annular kernels (see Fig.\,\ref{fig:Fig4}a1), defined by a common center $(i_0, j_0)$ and variable radii $r$. In our implementation, the ring radii are integer values from $r\,{=}\,1$ to $7$ pixels, with a width of $1$ pixel, i.\,e.\, each ring collects photons from pixels at distance $[r-0.5,r+0.5)$ from the center; the innermost kernel ($r\,{=}\,1$) is a filled disk of radius 1.5 pixels. Consequently, the measured number of photons on each pixel is now a vector $\vec{N}_m(r)$, with a single value for each kernel radius $r$. Taking advantage of the radial symmetry of the kernels and of the PSF, the atomic positions are entirely defined by the distance $d_a$ between the atom and the kernel center; see Fig.\,\ref{fig:Fig4}a3. For each kernel center, we then define the atomic configurations $C_\# = \{d_a\}_{a=1,...,n}$ by the set of distances of $n$ atoms. In practice, we build configurations placing up to 6 atoms at integer pixel radii, with a maximum distance $d_a\,{=}\,6$ pixels. The algorithmic implementation of Eq.\,\ref{eq:PNK} becomes:
\begin{equation}
    P(N \,|\, K) = P(N \,|\, C_\#=\{d_a\}_{a=1,...,n} \,, K(i_0,j_0, r)).
    \label{eq:PNK_algo}
\end{equation}
In Eq.\,\ref{eq:PNK_algo}, we explicitly define the configuration space over which Eq.\,\ref{eq:PNK} is evaluated. The radial symmetry allows to generate all possible configurations with a reasonably high number of atoms while avoiding combinatorial explosion. The local probability of any configuration $C_\#$ is then obtained from 
\begin{multline}
    P[C_\# \,|\, \vec{N}_m \, , (i_0, j_0)]\\
    = \frac{1}{\mathcal{N}} \prod_r P( C_\# \,|\, N = N_m(r) \,, K(i_0,j_0, r)),
\end{multline}
where $\mathcal{N}$ is a normalization factor ensuring $\sum_{C_\#}P[C_\# \,|\, \vec{N}_m \,|\, (i_0, j_0)]\,{=}\, 1$. This result holds with the statistical independence of the ring kernels: we ensure this with a zero overlap between the rings. Note that on the right-hand side, the conditional probabilities on the configurations have been inverted to conditional probabilities on the measured photon number. This is the result of a Bayes inversion with a flat prior. The local expectation value of the atom number at distance $d$ from the kernel center $(i_0, j_0)$ can then be calculated from the average over the atoms of each configuration:
\begin{equation}
    \langle n\,|\,(i_0,j_0)\rangle(d) = \sum_{C_\#} \sum_{d_a \in C_\#} P[C_\# \,|\, (i_0, j_0)] \times \delta(d-d_a).
\end{equation}
Note that this quantity is defined within the radial symmetry of the kernel center $(i_0,j_0)$, and the conditioning on $\vec{N}_m$ is implicit. Crucially, the choice of the radial symmetry inherently enables multilateration of the atom positions on the two-dimensional image grid; see Fig.\,\ref{fig:Fig4}b. Indeed, the average of $\langle n\,|\,(i_0,j_0)\rangle(d)$ over multiple kernel centers can be directly encoded on the pixels $(i,j)$ with:
\begin{equation}
    \langle n\rangle (i, j) = \frac{1}{\mathcal{N}} \sum_{(i_0,j_0)}
    \mathcal{D} \times w(d) \times \langle n\,|\,(i_0,j_0)\rangle (d),
    \label{eq:triangulation}
\end{equation}
where the distance criterion
\begin{equation}
    \mathcal{D}[(i,j), (i_0,j_0), d]\,{=}\,\delta( \norm{(i_0,j_0)-(i,j)}- d),
\end{equation}
triggers the conversion from radial distances to two-dimensional coordinates; $w(d)\,{=}\, 1/2\pi d$ is a geometric weight for the number of pixels at distance $d$ and $\mathcal{N}\,{=}\, \sum_{(i_0,j_0)} \mathcal{D}\times w(d)$  is a normalization factor. The result of the integration $\langle n\rangle(i, j)$ is then the expectation value of the atom number at pixel $(i,j)$. Note that a convenient consequence of this deconvolution process is the suppression of the background noise, as well as the photon shot-noise of each atom.\\

\begin{figure}[t!]
\centering    
\includegraphics[width=1\columnwidth]{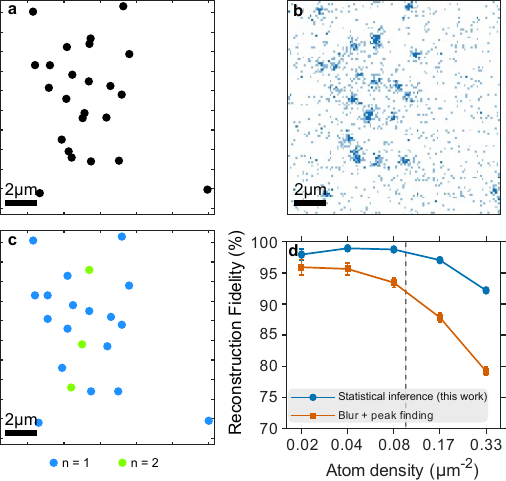}
\vspace{-0.0cm}
    \caption{\textbf{Monte-Carlo imaging simulations.} \textbf{a} Initial, random atomic positions. \textbf{b} The resulting image after Monte-Carlo simulation, $\tau\,{=}\,4\,\mathrm{\mu s}$. \textbf{c} Extracted atomic positions after statistical inference. \textbf{d} Reconstruction fidelity of both methods described in the main text, extracted from the imaging simulation, varying atomic density. Errorbars show the binomial standard error. The dashed line represents the atomic density of the image in Fig.\,\ref{fig:Fig4}.}
    \label{fig:SupMatMC}
\end{figure}

To convert the $\langle n\rangle (i, j)$ map into the most likely measured atomic configuration, i.\,e.\,a set of coordinates, we define a new set of configurations $\tilde{C}_\# \,{=}\, \{(i,j)^{\times m(i,j)}\}$. It is built from candidate positions, pinpointed by local maxima $\{(i, j)\}$ of $\langle n\rangle (i,j)$, and the multiplicity $m(i,j)$ can be constrained by $\langle n\rangle (i, j)$: the configuration space, in two-dimensional coordinates, is reduced from combinatorially large to tractable, without introducing any additional assumptions. The likelihood of every $\tilde{C}_\#$ is then evaluated with the same model on the raw image data, independently of the multilateration map; see Fig.\,\ref{fig:Fig4}c. To account for the cross-talk between neighboring atoms, peaks separated by less than the PSF extent are grouped into clusters and evaluated simultaneously. For each cluster, the joint likelihood of a configuration $\tilde{C}_\#$ is obtained by multiplying the local likelihoods over all kernel centers placed at the peak positions within the cluster:
\begin{equation}
    \Lambda(\tilde{C}_\#) = \prod_{(i_0,j_0)} P[\{ d = \norm{(i-i_0,j-j_0)}\} \,|\, \vec{N}_m \, , (i_0, j_0)].
    \label{eq:bayes_inference_atomNumber}
\end{equation}
The most likely configuration $\arg\max_{\tilde{C}_\#} \Lambda(\tilde{C}_\#)$ gives the final atom positions and multiplicities. When the kernel centers are separated by less than the diameter of the largest ring, the kernels overlap, and the conditional independence assumption underlying the likelihood product is violated. In this regime, the posterior is overconfident, but the evaluation of the most-likely configuration remains correct, as verified on simulated data; see Fig.\,\ref{fig:SupMatMC}. The rigorous treatment of dense clusters will be the subject of future work, e.\,g.\,with a more appropriate choice of kernel.

\section{\label{app:montecarlo} Algorithm benchmarking}

\noindent To characterize the efficiency of the atom detection algorithm, we perform a Monte-Carlo simulation of the imaging process that produces realistic images with known ground-truth atomic positions. Initial positions are drawn uniformly within the imaging plane and from a Gaussian distribution along $z$ with a width set by the pre-imaging harmonic oscillator length; velocities are sampled from a Maxwell-Boltzmann distribution at $20\,\mathrm{nK}$. See Fig.\,\ref{fig:SupMatMC}a for an exemplary atomic distribution.

The simulation mirrors the experimental sequence: counter-propagating imaging beams ($I\,{=}\, 30\,I_\mathrm{sat}$) are alternately pulsed; these beams are horizontally polarized in the lab frame. This pumps the $\sigma^{+/-}$ transitions of the atoms and produces a spontaneous emission pattern in which $97\%$ of the photons carry $\sigma^{+/-}$ polarization. Each absorbed and emitted photon imparts a recoil kick on the atom. For photons whose emission direction falls within the collection cone of the objective, detection occurs with a probability determined by the optical transmissions and quantum efficiencies along the imaging path. When a photon is registered on the simulated camera, its position on the sensor is obtained by projecting the current position of the atom onto the camera plane, offset by a random distance sampled from the experimental PSF, and rounding to the nearest pixel. This procedure captures atomic diffusion in the imaging plane. 
After imaging, noise drawn from the experimentally measured noise distribution of the camera is added to each image to simulate readout noise. Figure\,\ref{fig:SupMatMC}b shows the resultant image of the atomic distribution in a.

The resulting images are processed by either blurring and peak finding, or by the statistical deconvolution. Figure\,\ref{fig:SupMatMC}c shows the positions of atoms detected in b by the statistical inference algorithm. An atom is considered correctly identified if the deconvolution algorithm detects an atom within 1 FWHM of the atom’s initial position prior to imaging. Figure\,\ref{fig:SupMatMC}d shows how the fraction of correctly identified atoms scales with density for both detection methods.

\begin{figure}[t!]
\centering    
\includegraphics[width=1\columnwidth]{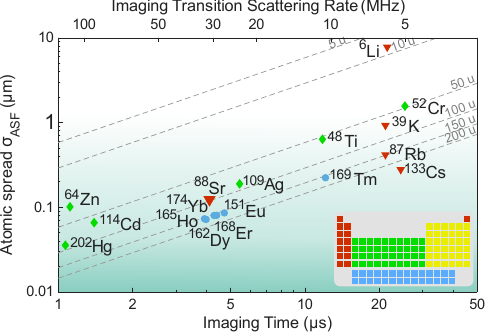}
\vspace{-0.25cm}
    \caption{\textbf{Feasibility of ultrafast imaging of various atomic species.} The in-plane RMS atomic spread around its time-averaged center for multiple laser-cooled atomic species, obtained via Monte-Carlo simulation. Imaging time is adjusted inversely proportional to the imaging transition linewidth. Dashed lines illustrate the mass and linewidth scaling for atoms with a transition at $450\,\unit{nm}$, obtained from Eq.\,\ref{eq:app_sigmaASF_2}. Parameters used for the calculation: $\eta_{\rm xy}\,{=}\, 0.8$ is the average of $\eta_{\rm x}$ and $\eta_{\rm y}$, $\epsilon \,{=}\, 10\%$, $N_{\rm det} \,{=}\, 40$ photons.}
    \label{fig:SupMatDiffSpecies}
\end{figure}

\section{\label{app:diffSpecies} Ultrafast Imaging of Various Species}

\noindent Our results are broadly applicable in the field of ultracold gases, as many atomic species can be imaged on microsecond timescales, maintaining near-diffraction limited resolution as a result of minimal atomic spread. Therefore, the relevant parameter to quantify the applicability of our method to any particular species is the atomic diffusion during the imaging readout, that we define as the atomic spread function $\sigma_\mathrm{ASF}$:
\begin{equation}
    \sigma_{\rm ASF} = \frac{h}{\Gamma m \lambda} \sqrt{\frac{2\eta_{\rm i} N_{\rm det}^3}{15\epsilon ^3}}
    \label{eq:app_sigmaASF_2}
\end{equation}
where, $N_{\rm det}$ is the number of detected photons by the camera and $\epsilon$ is the end-to-end efficiency of the imaging system. Note that $\sigma_\mathrm{ASF}$ is the RMS spread of the atom around its time-averaged position: $\sigma_{\rm ASF}(\tau) \,{=}\, \left < \frac{1}{\tau}\int_0^\tau(x(t)-\bar x)^2 dt \right >$, where $\tau \,{=}\, 2N_{\rm det}/\epsilon\Gamma$. To objectively compare $\sigma_{\rm ASF}$ between species, the signal on the camera, $N_{\rm det}$, is held constant. Here, $\eta_{\rm i}$ accounts for the anisotropy of the variance in velocity gained by absorption-emission cycles, where $i$ is the considered direction, $i\in\left\{x,y,z\right\}$. Each photon recoil adds a variance $\left< v_i^2\right> \,{=}\, \eta _i v_r^2$ along direction $i$. Absorption is only possible along the beam direction, while emission is possible in all directions, albeit with anisotropic probabilities when considering unbalanced ratios of $\sigma$ and $\pi$ polarized emitted light. In the experiment, we have counter-propagating beams along the $\pm x$ directions, and the emitted light is $\sigma ^{+/-}$ polarized, with the quantization axis pointing out of the imaging plane. This yields $\vec\eta \,{=}\,\left(\eta_x,\eta_y,\eta_z\right) \,{=}\, \left(1.3,0.3,0.4\right)$. The broadening in the perpendicular direction to the imaging beams is therefore a factor 2 smaller than the broadening parallel to the beams.

Figure\,\ref{fig:SupMatDiffSpecies} shows the applicability of ultrafast microscopy for 17 laser-cooled atomic species. The atomic spread function of each species was obtained from Monte-Carlo simulations. We verified that Eq.\,\ref{eq:app_sigmaASF_2} reproduces the simulated results, and therefore provides a reliable analytical expression for this figure of merit. Eleven of these species exhibit an RMS spread below $200\,\unit{nm}$, which we identify as particularly promising candidates for ultrafast imaging. Lanthanides and group-12 elements are of special interest, as their combination of a broad linewidth and a large mass is ideal for ultrafast imaging. Most notably, $^{202}$Hg with a transition at $185\,\unit{nm}$ and a linewidth of $\Gamma\,{=}\,2\pi \times 119\,\unit{MHz}$ is ideally suited for high-resolution ultrafast imaging, although this transition has yet to be investigated due to the technical challenges of deep UV generations \cite{lavigne2022mot}.

Ultrafast quantum gas microscopy is attractive for atomic species widely used in quantum simulation and quantum computation, such as ytterbium and strontium. Since the PSF and the ASF add in quadrature, the broadening of the fluorescence signal is only of second order in $\sigma_\mathrm{ASF}/\sigma_\mathrm{PSF}$. For a standard quantum gas microscope with $\mathrm{NA}\,{=}\,0.65$ and an end-to-end imaging efficiency of $10\,\%$, the fluorescence spot of ytterbium broadens by only about $18\,\%$, and that of strontium by about $35\,\%$, such that the resolution remains largely set by the imaging optics. 


\begin{thebibliography}{48}%
\makeatletter
\providecommand \@ifxundefined [1]{%
 \@ifx{#1\undefined}
}%
\providecommand \@ifnum [1]{%
 \ifnum #1\expandafter \@firstoftwo
 \else \expandafter \@secondoftwo
 \fi
}%
\providecommand \@ifx [1]{%
 \ifx #1\expandafter \@firstoftwo
 \else \expandafter \@secondoftwo
 \fi
}%
\providecommand \natexlab [1]{#1}%
\providecommand \enquote  [1]{``#1''}%
\providecommand \bibnamefont  [1]{#1}%
\providecommand \bibfnamefont [1]{#1}%
\providecommand \citenamefont [1]{#1}%
\providecommand \href@noop [0]{\@secondoftwo}%
\providecommand \href [0]{\begingroup \@sanitize@url \@href}%
\providecommand \@href[1]{\@@startlink{#1}\@@href}%
\providecommand \@@href[1]{\endgroup#1\@@endlink}%
\providecommand \@sanitize@url [0]{\catcode `\\12\catcode `\$12\catcode `\&12\catcode `\#12\catcode `\^12\catcode `\_12\catcode `\%12\relax}%
\providecommand \@@startlink[1]{}%
\providecommand \@@endlink[0]{}%
\providecommand \url  [0]{\begingroup\@sanitize@url \@url }%
\providecommand \@url [1]{\endgroup\@href {#1}{\urlprefix }}%
\providecommand \urlprefix  [0]{URL }%
\providecommand \Eprint [0]{\href }%
\providecommand \doibase [0]{https://doi.org/}%
\providecommand \selectlanguage [0]{\@gobble}%
\providecommand \bibinfo  [0]{\@secondoftwo}%
\providecommand \bibfield  [0]{\@secondoftwo}%
\providecommand \translation [1]{[#1]}%
\providecommand \BibitemOpen [0]{}%
\providecommand \bibitemStop [0]{}%
\providecommand \bibitemNoStop [0]{.\EOS\space}%
\providecommand \EOS [0]{\spacefactor3000\relax}%
\providecommand \BibitemShut  [1]{\csname bibitem#1\endcsname}%
\let\auto@bib@innerbib\@empty
\bibitem [{\citenamefont {Bakr}\ \emph {et~al.}(2009)\citenamefont {Bakr}, \citenamefont {Gillen}, \citenamefont {Peng}, \citenamefont {F{\"o}lling},\ and\ \citenamefont {Greiner}}]{Bakr2009aqg}%
  \BibitemOpen
  \bibfield  {author} {\bibinfo {author} {\bibfnamefont {W.~S.}\ \bibnamefont {Bakr}}, \bibinfo {author} {\bibfnamefont {J.~I.}\ \bibnamefont {Gillen}}, \bibinfo {author} {\bibfnamefont {A.}~\bibnamefont {Peng}}, \bibinfo {author} {\bibfnamefont {S.}~\bibnamefont {F{\"o}lling}},\ and\ \bibinfo {author} {\bibfnamefont {M.}~\bibnamefont {Greiner}},\ }\bibfield  {title} {\bibinfo {title} {{A quantum gas microscope for detecting single atoms in a Hubbard-regime optical lattice}},\ }\href {https://doi.org/10.1038/nature08482} {\bibfield  {journal} {\bibinfo  {journal} {Nature}\ }\textbf {\bibinfo {volume} {462}},\ \bibinfo {pages} {74} (\bibinfo {year} {2009})}\BibitemShut {NoStop}%
\bibitem [{\citenamefont {Sherson}\ \emph {et~al.}(2010)\citenamefont {Sherson}, \citenamefont {Weitenberg}, \citenamefont {Endres}, \citenamefont {Cheneau}, \citenamefont {Bloch},\ and\ \citenamefont {Kuhr}}]{Sherson2010sar}%
  \BibitemOpen
  \bibfield  {author} {\bibinfo {author} {\bibfnamefont {J.~F.}\ \bibnamefont {Sherson}}, \bibinfo {author} {\bibfnamefont {C.}~\bibnamefont {Weitenberg}}, \bibinfo {author} {\bibfnamefont {M.}~\bibnamefont {Endres}}, \bibinfo {author} {\bibfnamefont {M.}~\bibnamefont {Cheneau}}, \bibinfo {author} {\bibfnamefont {I.}~\bibnamefont {Bloch}},\ and\ \bibinfo {author} {\bibfnamefont {S.}~\bibnamefont {Kuhr}},\ }\bibfield  {title} {\bibinfo {title} {{Single-atom-resolved fluorescence imaging of an atomic Mott insulator}},\ }\href {https://doi.org/10.1038/nature09378} {\bibfield  {journal} {\bibinfo  {journal} {Nature}\ }\textbf {\bibinfo {volume} {467}},\ \bibinfo {pages} {68} (\bibinfo {year} {2010})}\BibitemShut {NoStop}%
\bibitem [{\citenamefont {Endres}\ \emph {et~al.}(2011)\citenamefont {Endres}, \citenamefont {Cheneau}, \citenamefont {Fukuhara}, \citenamefont {Weitenberg}, \citenamefont {Schauss}, \citenamefont {Gross}, \citenamefont {Mazza}, \citenamefont {Bañuls}, \citenamefont {Pollet}, \citenamefont {Bloch},\ and\ \citenamefont {Kuhr}}]{Endres2011ooc}%
  \BibitemOpen
  \bibfield  {author} {\bibinfo {author} {\bibfnamefont {M.}~\bibnamefont {Endres}}, \bibinfo {author} {\bibfnamefont {M.}~\bibnamefont {Cheneau}}, \bibinfo {author} {\bibfnamefont {T.}~\bibnamefont {Fukuhara}}, \bibinfo {author} {\bibfnamefont {C.}~\bibnamefont {Weitenberg}}, \bibinfo {author} {\bibfnamefont {P.}~\bibnamefont {Schauss}}, \bibinfo {author} {\bibfnamefont {C.}~\bibnamefont {Gross}}, \bibinfo {author} {\bibfnamefont {L.}~\bibnamefont {Mazza}}, \bibinfo {author} {\bibfnamefont {M.~C.}\ \bibnamefont {Bañuls}}, \bibinfo {author} {\bibfnamefont {L.}~\bibnamefont {Pollet}}, \bibinfo {author} {\bibfnamefont {I.}~\bibnamefont {Bloch}},\ and\ \bibinfo {author} {\bibfnamefont {S.}~\bibnamefont {Kuhr}},\ }\bibfield  {title} {\bibinfo {title} {{Observation of Correlated Particle-Hole Pairs and String Order in Low-Dimensional Mott Insulators.}},\ }\href {https://doi.org/10.1126/science.1209284} {\bibfield  {journal} {\bibinfo  {journal} {Science}\ }\textbf {\bibinfo {volume} {334}},\ \bibinfo {pages}
  {200} (\bibinfo {year} {2011})}\BibitemShut {NoStop}%
\bibitem [{\citenamefont {Mazurenko}\ \emph {et~al.}(2017)\citenamefont {Mazurenko}, \citenamefont {Chiu}, \citenamefont {Ji}, \citenamefont {Parsons}, \citenamefont {Kanász-Nagy}, \citenamefont {Schmidt}, \citenamefont {Grusdt}, \citenamefont {Demler}, \citenamefont {Greif},\ and\ \citenamefont {Greiner}}]{Mazurenko2017aca}%
  \BibitemOpen
  \bibfield  {author} {\bibinfo {author} {\bibfnamefont {A.}~\bibnamefont {Mazurenko}}, \bibinfo {author} {\bibfnamefont {C.~S.}\ \bibnamefont {Chiu}}, \bibinfo {author} {\bibfnamefont {G.}~\bibnamefont {Ji}}, \bibinfo {author} {\bibfnamefont {M.~F.}\ \bibnamefont {Parsons}}, \bibinfo {author} {\bibfnamefont {M.}~\bibnamefont {Kanász-Nagy}}, \bibinfo {author} {\bibfnamefont {R.}~\bibnamefont {Schmidt}}, \bibinfo {author} {\bibfnamefont {F.}~\bibnamefont {Grusdt}}, \bibinfo {author} {\bibfnamefont {E.}~\bibnamefont {Demler}}, \bibinfo {author} {\bibfnamefont {D.}~\bibnamefont {Greif}},\ and\ \bibinfo {author} {\bibfnamefont {M.}~\bibnamefont {Greiner}},\ }\bibfield  {title} {\bibinfo {title} {{A cold-atom Fermi--Hubbard antiferromagnet}},\ }\href {https://doi.org/10.1038/nature22362} {\bibfield  {journal} {\bibinfo  {journal} {Nature}\ }\textbf {\bibinfo {volume} {545}},\ \bibinfo {pages} {462} (\bibinfo {year} {2017})}\BibitemShut {NoStop}%
\bibitem [{\citenamefont {Hartke}\ \emph {et~al.}(2023)\citenamefont {Hartke}, \citenamefont {Oreg}, \citenamefont {Turnbaugh}, \citenamefont {Jia},\ and\ \citenamefont {Zwierlein}}]{Hartke2023doo}%
  \BibitemOpen
  \bibfield  {author} {\bibinfo {author} {\bibfnamefont {T.}~\bibnamefont {Hartke}}, \bibinfo {author} {\bibfnamefont {B.}~\bibnamefont {Oreg}}, \bibinfo {author} {\bibfnamefont {C.}~\bibnamefont {Turnbaugh}}, \bibinfo {author} {\bibfnamefont {N.}~\bibnamefont {Jia}},\ and\ \bibinfo {author} {\bibfnamefont {M.}~\bibnamefont {Zwierlein}},\ }\bibfield  {title} {\bibinfo {title} {Direct observation of nonlocal fermion pairing in an attractive {Fermi-Hubbard} gas.},\ }\href {https://doi.org/10.1126/science.ade4245} {\bibfield  {journal} {\bibinfo  {journal} {Science}\ }\textbf {\bibinfo {volume} {381}},\ \bibinfo {pages} {82} (\bibinfo {year} {2023})}\BibitemShut {NoStop}%
\bibitem [{\citenamefont {Gross}\ and\ \citenamefont {Bakr}(2021)}]{Gross2021qgm}%
  \BibitemOpen
  \bibfield  {author} {\bibinfo {author} {\bibfnamefont {C.}~\bibnamefont {Gross}}\ and\ \bibinfo {author} {\bibfnamefont {W.~S.}\ \bibnamefont {Bakr}},\ }\bibfield  {title} {\bibinfo {title} {Quantum gas microscopy for single atom and spin detection},\ }\href {https://doi.org/10.1038/s41567-021-01370-5} {\bibfield  {journal} {\bibinfo  {journal} {Nature Physics}\ }\textbf {\bibinfo {volume} {17}},\ \bibinfo {pages} {1316} (\bibinfo {year} {2021})}\BibitemShut {NoStop}%
\bibitem [{\citenamefont {Kaufman}\ and\ \citenamefont {Ni}(2021)}]{Kaufman2021qsw}%
  \BibitemOpen
  \bibfield  {author} {\bibinfo {author} {\bibfnamefont {A.}~\bibnamefont {Kaufman}}\ and\ \bibinfo {author} {\bibfnamefont {K.-K.}\ \bibnamefont {Ni}},\ }\bibfield  {title} {\bibinfo {title} {Quantum science with optical tweezer arrays of ultracold atoms and molecules.},\ }\href {https://doi.org/https://doi.org/10.1038/s41567-021-01357-2} {\bibfield  {journal} {\bibinfo  {journal} {Nat. Phys.}\ }\textbf {\bibinfo {volume} {17}},\ \bibinfo {pages} {1324–1333} (\bibinfo {year} {2021})}\BibitemShut {NoStop}%
\bibitem [{\citenamefont {Ott}\ \emph {et~al.}(2024)\citenamefont {Ott}, \citenamefont {Zache}, \citenamefont {Pr\"{u}fer}, \citenamefont {Erne}, \citenamefont {Tajik}, \citenamefont {Pichler}, \citenamefont {Schmiedmayer},\ and\ \citenamefont {Zoller}}]{Ott2024hlq}%
  \BibitemOpen
  \bibfield  {author} {\bibinfo {author} {\bibfnamefont {R.}~\bibnamefont {Ott}}, \bibinfo {author} {\bibfnamefont {T.~V.}\ \bibnamefont {Zache}}, \bibinfo {author} {\bibfnamefont {M.}~\bibnamefont {Pr\"{u}fer}}, \bibinfo {author} {\bibfnamefont {S.}~\bibnamefont {Erne}}, \bibinfo {author} {\bibfnamefont {M.}~\bibnamefont {Tajik}}, \bibinfo {author} {\bibfnamefont {H.}~\bibnamefont {Pichler}}, \bibinfo {author} {\bibfnamefont {J.}~\bibnamefont {Schmiedmayer}},\ and\ \bibinfo {author} {\bibfnamefont {P.}~\bibnamefont {Zoller}},\ }\bibfield  {title} {\bibinfo {title} {Hamiltonian learning in quantum field theories},\ }\href {https://doi.org/10.1103/PhysRevResearch.6.043284} {\bibfield  {journal} {\bibinfo  {journal} {Physical Review Research}\ }\textbf {\bibinfo {volume} {6}},\ \bibinfo {pages} {043284} (\bibinfo {year} {2024})}\BibitemShut {NoStop}%
\bibitem [{\citenamefont {Esposito}\ \emph {et~al.}(2009)\citenamefont {Esposito}, \citenamefont {Harbola},\ and\ \citenamefont {Mukamel}}]{Esposito2009nff}%
  \BibitemOpen
  \bibfield  {author} {\bibinfo {author} {\bibfnamefont {M.}~\bibnamefont {Esposito}}, \bibinfo {author} {\bibfnamefont {U.}~\bibnamefont {Harbola}},\ and\ \bibinfo {author} {\bibfnamefont {S.}~\bibnamefont {Mukamel}},\ }\bibfield  {title} {\bibinfo {title} {Nonequilibrium fluctuations, fluctuation theorems, and counting statistics in quantum systems},\ }\href {https://doi.org/10.1103/RevModPhys.81.1665} {\bibfield  {journal} {\bibinfo  {journal} {Rev. Mod. Phys.}\ }\textbf {\bibinfo {volume} {81}},\ \bibinfo {pages} {1665} (\bibinfo {year} {2009})}\BibitemShut {NoStop}%
\bibitem [{\citenamefont {Polkovnikov}\ \emph {et~al.}(2011)\citenamefont {Polkovnikov}, \citenamefont {Sengupta}, \citenamefont {Silva},\ and\ \citenamefont {Vengalattore}}]{Polkovnikov2011ndo}%
  \BibitemOpen
  \bibfield  {author} {\bibinfo {author} {\bibfnamefont {A.}~\bibnamefont {Polkovnikov}}, \bibinfo {author} {\bibfnamefont {K.}~\bibnamefont {Sengupta}}, \bibinfo {author} {\bibfnamefont {A.}~\bibnamefont {Silva}},\ and\ \bibinfo {author} {\bibfnamefont {M.}~\bibnamefont {Vengalattore}},\ }\bibfield  {title} {\bibinfo {title} {Colloquium: Nonequilibrium dynamics of closed interacting quantum systems},\ }\href {https://doi.org/10.1103/RevModPhys.83.863} {\bibfield  {journal} {\bibinfo  {journal} {Rev. Mod. Phys.}\ }\textbf {\bibinfo {volume} {83}},\ \bibinfo {pages} {863} (\bibinfo {year} {2011})}\BibitemShut {NoStop}%
\bibitem [{\citenamefont {Eisert}\ \emph {et~al.}(2015)\citenamefont {Eisert}, \citenamefont {Friesdorf},\ and\ \citenamefont {Gogolin}}]{Eisert2015qmb}%
  \BibitemOpen
  \bibfield  {author} {\bibinfo {author} {\bibfnamefont {J.}~\bibnamefont {Eisert}}, \bibinfo {author} {\bibfnamefont {M.}~\bibnamefont {Friesdorf}},\ and\ \bibinfo {author} {\bibfnamefont {C.}~\bibnamefont {Gogolin}},\ }\bibfield  {title} {\bibinfo {title} {{Quantum many-body systems out of equilibrium}},\ }\href {https://doi.org/10.1038/nphys3215} {\bibfield  {journal} {\bibinfo  {journal} {Nature Phys.}\ }\textbf {\bibinfo {volume} {11}},\ \bibinfo {pages} {124} (\bibinfo {year} {2015})}\BibitemShut {NoStop}%
\bibitem [{\citenamefont {Chalopin}\ \emph {et~al.}(2026)\citenamefont {Chalopin}, \citenamefont {Ferrier-Barbut}, \citenamefont {Lahaye}, \citenamefont {Browaeys},\ and\ \citenamefont {Cl\'ement}}]{Chalopin2026cci}%
  \BibitemOpen
  \bibfield  {author} {\bibinfo {author} {\bibfnamefont {T.}~\bibnamefont {Chalopin}}, \bibinfo {author} {\bibfnamefont {I.}~\bibnamefont {Ferrier-Barbut}}, \bibinfo {author} {\bibfnamefont {T.}~\bibnamefont {Lahaye}}, \bibinfo {author} {\bibfnamefont {A.}~\bibnamefont {Browaeys}},\ and\ \bibinfo {author} {\bibfnamefont {D.}~\bibnamefont {Cl\'ement}},\ }\bibfield  {title} {\bibinfo {title} {Connected correlations in cold atom experiments},\ }\href {https://doi.org/10.5802/crphys.274} {\bibfield  {journal} {\bibinfo  {journal} {Comptes Rendus. Physique}\ }\textbf {\bibinfo {volume} {27}},\ \bibinfo {pages} {65} (\bibinfo {year} {2026})}\BibitemShut {NoStop}%
\bibitem [{\citenamefont {Dixmerias}\ \emph {et~al.}(2026{\natexlab{a}})\citenamefont {Dixmerias}, \citenamefont {Verstraten}, \citenamefont {Daix}, \citenamefont {Peaudecerf}, \citenamefont {de~Jongh},\ and\ \citenamefont {Yefsah}}]{Dixmerias2026ftq}%
  \BibitemOpen
  \bibfield  {author} {\bibinfo {author} {\bibfnamefont {M.}~\bibnamefont {Dixmerias}}, \bibinfo {author} {\bibfnamefont {J.}~\bibnamefont {Verstraten}}, \bibinfo {author} {\bibfnamefont {C.}~\bibnamefont {Daix}}, \bibinfo {author} {\bibfnamefont {B.}~\bibnamefont {Peaudecerf}}, \bibinfo {author} {\bibfnamefont {T.}~\bibnamefont {de~Jongh}},\ and\ \bibinfo {author} {\bibfnamefont {T.}~\bibnamefont {Yefsah}},\ }\bibfield  {title} {\bibinfo {title} {Fluctuation thermometry of an atom-resolved quantum gas: Beyond the fluctuation-dissipation theorem},\ }\href {https://doi.org/10.1103/45fw-r1cp} {\bibfield  {journal} {\bibinfo  {journal} {Phys. Rev. Lett.}\ }\textbf {\bibinfo {volume} {137}},\ \bibinfo {pages} {123401} (\bibinfo {year} {2026}{\natexlab{a}})}\BibitemShut {NoStop}%
\bibitem [{\citenamefont {Xiang}\ \emph {et~al.}(2025)\citenamefont {Xiang}, \citenamefont {Cruz-Col\'on}, \citenamefont {Chua}, \citenamefont {Milner}, \citenamefont {de~Hond}, \citenamefont {Fricke},\ and\ \citenamefont {Ketterle}}]{Xiang2025isi}%
  \BibitemOpen
  \bibfield  {author} {\bibinfo {author} {\bibfnamefont {J.}~\bibnamefont {Xiang}}, \bibinfo {author} {\bibfnamefont {E.}~\bibnamefont {Cruz-Col\'on}}, \bibinfo {author} {\bibfnamefont {C.~C.}\ \bibnamefont {Chua}}, \bibinfo {author} {\bibfnamefont {W.~R.}\ \bibnamefont {Milner}}, \bibinfo {author} {\bibfnamefont {J.}~\bibnamefont {de~Hond}}, \bibinfo {author} {\bibfnamefont {J.~F.}\ \bibnamefont {Fricke}},\ and\ \bibinfo {author} {\bibfnamefont {W.}~\bibnamefont {Ketterle}},\ }\bibfield  {title} {\bibinfo {title} {{\textit{In Situ} Imaging of the Thermal de Broglie Wavelength in an Ultracold Bose Gas}},\ }\href {https://doi.org/10.1103/PhysRevLett.134.183401} {\bibfield  {journal} {\bibinfo  {journal} {Phys. Rev. Lett.}\ }\textbf {\bibinfo {volume} {134}},\ \bibinfo {pages} {183401} (\bibinfo {year} {2025})}\BibitemShut {NoStop}%
\bibitem [{\citenamefont {Yao}\ \emph {et~al.}(2025)\citenamefont {Yao}, \citenamefont {Chi}, \citenamefont {Wang}, \citenamefont {Fletcher},\ and\ \citenamefont {Zwierlein}}]{Yao2025mpc}%
  \BibitemOpen
  \bibfield  {author} {\bibinfo {author} {\bibfnamefont {R.}~\bibnamefont {Yao}}, \bibinfo {author} {\bibfnamefont {S.}~\bibnamefont {Chi}}, \bibinfo {author} {\bibfnamefont {M.}~\bibnamefont {Wang}}, \bibinfo {author} {\bibfnamefont {R.~J.}\ \bibnamefont {Fletcher}},\ and\ \bibinfo {author} {\bibfnamefont {M.}~\bibnamefont {Zwierlein}},\ }\bibfield  {title} {\bibinfo {title} {{Measuring Pair Correlations in Bose and Fermi Gases via Atom-Resolved Microscopy}},\ }\href {https://doi.org/10.1103/PhysRevLett.134.183402} {\bibfield  {journal} {\bibinfo  {journal} {Phys. Rev. Lett.}\ }\textbf {\bibinfo {volume} {134}},\ \bibinfo {pages} {183402} (\bibinfo {year} {2025})}\BibitemShut {NoStop}%
\bibitem [{\citenamefont {de~Jongh}\ \emph {et~al.}(2025)\citenamefont {de~Jongh}, \citenamefont {Verstraten}, \citenamefont {Dixmerias}, \citenamefont {Daix}, \citenamefont {Peaudecerf},\ and\ \citenamefont {Yefsah}}]{deJongh2025qgm}%
  \BibitemOpen
  \bibfield  {author} {\bibinfo {author} {\bibfnamefont {T.}~\bibnamefont {de~Jongh}}, \bibinfo {author} {\bibfnamefont {J.}~\bibnamefont {Verstraten}}, \bibinfo {author} {\bibfnamefont {M.}~\bibnamefont {Dixmerias}}, \bibinfo {author} {\bibfnamefont {C.}~\bibnamefont {Daix}}, \bibinfo {author} {\bibfnamefont {B.}~\bibnamefont {Peaudecerf}},\ and\ \bibinfo {author} {\bibfnamefont {T.}~\bibnamefont {Yefsah}},\ }\bibfield  {title} {\bibinfo {title} {{Quantum Gas Microscopy of Fermions in the Continuum}},\ }\href {https://doi.org/10.1103/PhysRevLett.134.183403} {\bibfield  {journal} {\bibinfo  {journal} {Phys. Rev. Lett.}\ }\textbf {\bibinfo {volume} {134}},\ \bibinfo {pages} {183403} (\bibinfo {year} {2025})}\BibitemShut {NoStop}%
\bibitem [{\citenamefont {Daix}\ \emph {et~al.}(2026)\citenamefont {Daix}, \citenamefont {Dixmerias}, \citenamefont {He}, \citenamefont {Verstraten}, \citenamefont {de~Jongh}, \citenamefont {Peaudecerf}, \citenamefont {Zhang},\ and\ \citenamefont {Yefsah}}]{Daix2026osc}%
  \BibitemOpen
  \bibfield  {author} {\bibinfo {author} {\bibfnamefont {C.}~\bibnamefont {Daix}}, \bibinfo {author} {\bibfnamefont {M.}~\bibnamefont {Dixmerias}}, \bibinfo {author} {\bibfnamefont {Y.-Y.}\ \bibnamefont {He}}, \bibinfo {author} {\bibfnamefont {J.}~\bibnamefont {Verstraten}}, \bibinfo {author} {\bibfnamefont {T.}~\bibnamefont {de~Jongh}}, \bibinfo {author} {\bibfnamefont {B.}~\bibnamefont {Peaudecerf}}, \bibinfo {author} {\bibfnamefont {S.}~\bibnamefont {Zhang}},\ and\ \bibinfo {author} {\bibfnamefont {T.}~\bibnamefont {Yefsah}},\ }\bibfield  {title} {\bibinfo {title} {Observing spatial charge and spin correlations in a strongly interacting {F}ermi gas},\ }\href {https://doi.org/10.1103/2t2k-3ftx} {\bibfield  {journal} {\bibinfo  {journal} {Phys. Rev. Lett.}\ }\textbf {\bibinfo {volume} {136}},\ \bibinfo {pages} {153402} (\bibinfo {year} {2026})}\BibitemShut {NoStop}%
\bibitem [{\citenamefont {Dixmerias}\ \emph {et~al.}(2026{\natexlab{b}})\citenamefont {Dixmerias}, \citenamefont {Del Vecchio Del~Vecchio}, \citenamefont {Daix}, \citenamefont {Verstraten}, \citenamefont {de~Jongh}, \citenamefont {Peaudecerf}, \citenamefont {Le~Doussal}, \citenamefont {Schehr},\ and\ \citenamefont {Yefsah}}]{Dixmerias2026urm}%
  \BibitemOpen
  \bibfield  {author} {\bibinfo {author} {\bibfnamefont {M.}~\bibnamefont {Dixmerias}}, \bibinfo {author} {\bibfnamefont {G.}~\bibnamefont {Del Vecchio Del~Vecchio}}, \bibinfo {author} {\bibfnamefont {C.}~\bibnamefont {Daix}}, \bibinfo {author} {\bibfnamefont {J.}~\bibnamefont {Verstraten}}, \bibinfo {author} {\bibfnamefont {T.}~\bibnamefont {de~Jongh}}, \bibinfo {author} {\bibfnamefont {B.}~\bibnamefont {Peaudecerf}}, \bibinfo {author} {\bibfnamefont {P.}~\bibnamefont {Le~Doussal}}, \bibinfo {author} {\bibfnamefont {G.}~\bibnamefont {Schehr}},\ and\ \bibinfo {author} {\bibfnamefont {T.}~\bibnamefont {Yefsah}},\ }\bibfield  {title} {\bibinfo {title} {Universal random matrix behavior of a fermionic quantum gas},\ }\bibfield  {journal} {\bibinfo  {journal} {Nat. Commun.}\ }\href {https://doi.org/10.1038/s41467-026-77088-w} {10.1038/s41467-026-77088-w} (\bibinfo {year} {2026}{\natexlab{b}})\BibitemShut {NoStop}%
\bibitem [{\citenamefont {Bücker}\ \emph {et~al.}(2009)\citenamefont {Bücker}, \citenamefont {Perrin}, \citenamefont {Manz}, \citenamefont {Betz}, \citenamefont {Koller}, \citenamefont {Plisson}, \citenamefont {Rottmann}, \citenamefont {Schumm},\ and\ \citenamefont {Schmiedmayer}}]{Buecker2009sps}%
  \BibitemOpen
  \bibfield  {author} {\bibinfo {author} {\bibfnamefont {R.}~\bibnamefont {Bücker}}, \bibinfo {author} {\bibfnamefont {A.}~\bibnamefont {Perrin}}, \bibinfo {author} {\bibfnamefont {S.}~\bibnamefont {Manz}}, \bibinfo {author} {\bibfnamefont {T.}~\bibnamefont {Betz}}, \bibinfo {author} {\bibfnamefont {C.}~\bibnamefont {Koller}}, \bibinfo {author} {\bibfnamefont {T.}~\bibnamefont {Plisson}}, \bibinfo {author} {\bibfnamefont {J.}~\bibnamefont {Rottmann}}, \bibinfo {author} {\bibfnamefont {T.}~\bibnamefont {Schumm}},\ and\ \bibinfo {author} {\bibfnamefont {J.}~\bibnamefont {Schmiedmayer}},\ }\bibfield  {title} {\bibinfo {title} {{Single-particle-sensitive imaging of freely propagating ultracold atoms}},\ }\href {https://doi.org/10.1088/1367-2630/11/10/103039} {\bibfield  {journal} {\bibinfo  {journal} {New Journal of Physics}\ }\textbf {\bibinfo {volume} {11}},\ \bibinfo {pages} {103039} (\bibinfo {year} {2009})}\BibitemShut {NoStop}%
\bibitem [{\citenamefont {Fuhrmanek}\ \emph {et~al.}(2010)\citenamefont {Fuhrmanek}, \citenamefont {Lance}, \citenamefont {Tuchendler}, \citenamefont {Grangier}, \citenamefont {Sortais},\ and\ \citenamefont {Browaeys}}]{Fuhrmanek2010ias}%
  \BibitemOpen
  \bibfield  {author} {\bibinfo {author} {\bibfnamefont {A.}~\bibnamefont {Fuhrmanek}}, \bibinfo {author} {\bibfnamefont {A.~M.}\ \bibnamefont {Lance}}, \bibinfo {author} {\bibfnamefont {C.}~\bibnamefont {Tuchendler}}, \bibinfo {author} {\bibfnamefont {P.}~\bibnamefont {Grangier}}, \bibinfo {author} {\bibfnamefont {Y.~R.~P.}\ \bibnamefont {Sortais}},\ and\ \bibinfo {author} {\bibfnamefont {A.}~\bibnamefont {Browaeys}},\ }\bibfield  {title} {\bibinfo {title} {Imaging a single atom in a time-of-flight experiment},\ }\href {https://doi.org/10.1088/1367-2630/12/5/053028} {\bibfield  {journal} {\bibinfo  {journal} {New J. Phys.}\ }\textbf {\bibinfo {volume} {12}},\ \bibinfo {pages} {053028} (\bibinfo {year} {2010})}\BibitemShut {NoStop}%
\bibitem [{\citenamefont {Bergschneider}\ \emph {et~al.}(2018)\citenamefont {Bergschneider}, \citenamefont {Klinkhamer}, \citenamefont {Becher}, \citenamefont {Klemt}, \citenamefont {Z\"urn}, \citenamefont {Preiss},\ and\ \citenamefont {Jochim}}]{Bergschneider2018srs}%
  \BibitemOpen
  \bibfield  {author} {\bibinfo {author} {\bibfnamefont {A.}~\bibnamefont {Bergschneider}}, \bibinfo {author} {\bibfnamefont {V.~M.}\ \bibnamefont {Klinkhamer}}, \bibinfo {author} {\bibfnamefont {J.~H.}\ \bibnamefont {Becher}}, \bibinfo {author} {\bibfnamefont {R.}~\bibnamefont {Klemt}}, \bibinfo {author} {\bibfnamefont {G.}~\bibnamefont {Z\"urn}}, \bibinfo {author} {\bibfnamefont {P.~M.}\ \bibnamefont {Preiss}},\ and\ \bibinfo {author} {\bibfnamefont {S.}~\bibnamefont {Jochim}},\ }\bibfield  {title} {\bibinfo {title} {Spin-resolved single-atom imaging of $^{6}\mathrm{Li}$ in free space},\ }\href {https://doi.org/10.1103/PhysRevA.97.063613} {\bibfield  {journal} {\bibinfo  {journal} {Phys. Rev. A}\ }\textbf {\bibinfo {volume} {97}},\ \bibinfo {pages} {063613} (\bibinfo {year} {2018})}\BibitemShut {NoStop}%
\bibitem [{\citenamefont {Picard}\ \emph {et~al.}(2020)\citenamefont {Picard}, \citenamefont {Mark}, \citenamefont {Ferlaino},\ and\ \citenamefont {van Bijnen}}]{Picard2020dla}%
  \BibitemOpen
  \bibfield  {author} {\bibinfo {author} {\bibfnamefont {L.~R.~B.}\ \bibnamefont {Picard}}, \bibinfo {author} {\bibfnamefont {M.~J.}\ \bibnamefont {Mark}}, \bibinfo {author} {\bibfnamefont {F.}~\bibnamefont {Ferlaino}},\ and\ \bibinfo {author} {\bibfnamefont {R.}~\bibnamefont {van Bijnen}},\ }\bibfield  {title} {\bibinfo {title} {Deep learning-assisted classification of site-resolved quantum gas microscope images},\ }\href {https://doi.org/10.1088/1361-6501/ab44d8} {\bibfield  {journal} {\bibinfo  {journal} {Meas. Sci. Technol.}\ }\textbf {\bibinfo {volume} {31}},\ \bibinfo {pages} {025201} (\bibinfo {year} {2020})}\BibitemShut {NoStop}%
\bibitem [{\citenamefont {Su}\ \emph {et~al.}(2023)\citenamefont {Su}, \citenamefont {Douglas}, \citenamefont {Szurek}, \citenamefont {Groth}, \citenamefont {Ozturk}, \citenamefont {Krahn}, \citenamefont {H{\'e}bert}, \citenamefont {Phelps}, \citenamefont {Ebadi}, \citenamefont {Dickerson}, \citenamefont {Ferlaino}, \citenamefont {Markovi{\'{c}}},\ and\ \citenamefont {Greiner}}]{Su2023dqs}%
  \BibitemOpen
  \bibfield  {author} {\bibinfo {author} {\bibfnamefont {L.}~\bibnamefont {Su}}, \bibinfo {author} {\bibfnamefont {A.}~\bibnamefont {Douglas}}, \bibinfo {author} {\bibfnamefont {M.}~\bibnamefont {Szurek}}, \bibinfo {author} {\bibfnamefont {R.}~\bibnamefont {Groth}}, \bibinfo {author} {\bibfnamefont {S.~F.}\ \bibnamefont {Ozturk}}, \bibinfo {author} {\bibfnamefont {A.}~\bibnamefont {Krahn}}, \bibinfo {author} {\bibfnamefont {A.~H.}\ \bibnamefont {H{\'e}bert}}, \bibinfo {author} {\bibfnamefont {G.~A.}\ \bibnamefont {Phelps}}, \bibinfo {author} {\bibfnamefont {S.}~\bibnamefont {Ebadi}}, \bibinfo {author} {\bibfnamefont {S.}~\bibnamefont {Dickerson}}, \bibinfo {author} {\bibfnamefont {F.}~\bibnamefont {Ferlaino}}, \bibinfo {author} {\bibfnamefont {O.}~\bibnamefont {Markovi{\'{c}}}},\ and\ \bibinfo {author} {\bibfnamefont {M.}~\bibnamefont {Greiner}},\ }\bibfield  {title} {\bibinfo {title} {Dipolar quantum solids emerging in a {Hubbard} quantum simulator},\ }\href {https://doi.org/10.1038/s41586-023-06614-3}
  {\bibfield  {journal} {\bibinfo  {journal} {Nature}\ }\textbf {\bibinfo {volume} {622}},\ \bibinfo {pages} {724} (\bibinfo {year} {2023})}\BibitemShut {NoStop}%
\bibitem [{\citenamefont {Su}\ \emph {et~al.}(2025)\citenamefont {Su}, \citenamefont {Douglas}, \citenamefont {Szurek}, \citenamefont {Hebert}, \citenamefont {Krahn}, \citenamefont {Groth}, \citenamefont {Phelps}, \citenamefont {Markovic},\ and\ \citenamefont {Greiner}}]{Su2025fsa}%
  \BibitemOpen
  \bibfield  {author} {\bibinfo {author} {\bibfnamefont {L.}~\bibnamefont {Su}}, \bibinfo {author} {\bibfnamefont {A.}~\bibnamefont {Douglas}}, \bibinfo {author} {\bibfnamefont {M.}~\bibnamefont {Szurek}}, \bibinfo {author} {\bibfnamefont {A.~H.}\ \bibnamefont {Hebert}}, \bibinfo {author} {\bibfnamefont {A.}~\bibnamefont {Krahn}}, \bibinfo {author} {\bibfnamefont {R.}~\bibnamefont {Groth}}, \bibinfo {author} {\bibfnamefont {G.~A.}\ \bibnamefont {Phelps}}, \bibinfo {author} {\bibfnamefont {O.}~\bibnamefont {Markovic}},\ and\ \bibinfo {author} {\bibfnamefont {M.}~\bibnamefont {Greiner}},\ }\bibfield  {title} {\bibinfo {title} {Fast single atom imaging in optical lattice arrays},\ }\href {https://doi.org/10.1038/s41467-025-56305-y} {\bibfield  {journal} {\bibinfo  {journal} {Nature communications}\ }\textbf {\bibinfo {volume} {16}},\ \bibinfo {pages} {1017} (\bibinfo {year} {2025})}\BibitemShut {NoStop}%
\bibitem [{\citenamefont {Gr\"un}\ \emph {et~al.}(2024)\citenamefont {Gr\"un}, \citenamefont {White}, \citenamefont {Ortu}, \citenamefont {Di~Carli}, \citenamefont {Edri}, \citenamefont {Lepers}, \citenamefont {Mark},\ and\ \citenamefont {Ferlaino}}]{Gruen2024ota}%
  \BibitemOpen
  \bibfield  {author} {\bibinfo {author} {\bibfnamefont {D.~S.}\ \bibnamefont {Gr\"un}}, \bibinfo {author} {\bibfnamefont {S.~J.~M.}\ \bibnamefont {White}}, \bibinfo {author} {\bibfnamefont {A.}~\bibnamefont {Ortu}}, \bibinfo {author} {\bibfnamefont {A.}~\bibnamefont {Di~Carli}}, \bibinfo {author} {\bibfnamefont {H.}~\bibnamefont {Edri}}, \bibinfo {author} {\bibfnamefont {M.}~\bibnamefont {Lepers}}, \bibinfo {author} {\bibfnamefont {M.~J.}\ \bibnamefont {Mark}},\ and\ \bibinfo {author} {\bibfnamefont {F.}~\bibnamefont {Ferlaino}},\ }\bibfield  {title} {\bibinfo {title} {{Optical Tweezer Arrays of Erbium Atoms}},\ }\href {https://doi.org/10.1103/PhysRevLett.133.223402} {\bibfield  {journal} {\bibinfo  {journal} {Phys. Rev. Lett.}\ }\textbf {\bibinfo {volume} {133}},\ \bibinfo {pages} {223402} (\bibinfo {year} {2024})}\BibitemShut {NoStop}%
\bibitem [{\citenamefont {Gr\"un}\ \emph {et~al.}(2026)\citenamefont {Gr\"un}, \citenamefont {Giacomelli}, \citenamefont {Tashchilina}, \citenamefont {Donofrio}, \citenamefont {Borchers}, \citenamefont {Bland}, \citenamefont {Mark},\ and\ \citenamefont {Ferlaino}}]{Gruen2026lac}%
  \BibitemOpen
  \bibfield  {author} {\bibinfo {author} {\bibfnamefont {D.~S.}\ \bibnamefont {Gr\"un}}, \bibinfo {author} {\bibfnamefont {L.~B.}\ \bibnamefont {Giacomelli}}, \bibinfo {author} {\bibfnamefont {A.}~\bibnamefont {Tashchilina}}, \bibinfo {author} {\bibfnamefont {R.}~\bibnamefont {Donofrio}}, \bibinfo {author} {\bibfnamefont {F.}~\bibnamefont {Borchers}}, \bibinfo {author} {\bibfnamefont {T.}~\bibnamefont {Bland}}, \bibinfo {author} {\bibfnamefont {M.~J.}\ \bibnamefont {Mark}},\ and\ \bibinfo {author} {\bibfnamefont {F.}~\bibnamefont {Ferlaino}},\ }\bibfield  {title} {\bibinfo {title} {Light-assisted collisions in tweezer-trapped lanthanides},\ }\href {https://doi.org/10.1103/rkn3-8rc9} {\bibfield  {journal} {\bibinfo  {journal} {Phys. Rev. Res.}\ }\textbf {\bibinfo {volume} {8}},\ \bibinfo {pages} {023067} (\bibinfo {year} {2026})}\BibitemShut {NoStop}%
\bibitem [{\citenamefont {Ma}\ \emph {et~al.}(2023)\citenamefont {Ma}, \citenamefont {Liu}, \citenamefont {Peng}, \citenamefont {Zhang}, \citenamefont {Jandura}, \citenamefont {Claes}, \citenamefont {Burgers}, \citenamefont {Pupillo}, \citenamefont {Puri},\ and\ \citenamefont {Thompson}}]{Ma2023hfg}%
  \BibitemOpen
  \bibfield  {author} {\bibinfo {author} {\bibfnamefont {S.}~\bibnamefont {Ma}}, \bibinfo {author} {\bibfnamefont {G.}~\bibnamefont {Liu}}, \bibinfo {author} {\bibfnamefont {P.}~\bibnamefont {Peng}}, \bibinfo {author} {\bibfnamefont {B.}~\bibnamefont {Zhang}}, \bibinfo {author} {\bibfnamefont {S.}~\bibnamefont {Jandura}}, \bibinfo {author} {\bibfnamefont {J.}~\bibnamefont {Claes}}, \bibinfo {author} {\bibfnamefont {A.~P.}\ \bibnamefont {Burgers}}, \bibinfo {author} {\bibfnamefont {G.}~\bibnamefont {Pupillo}}, \bibinfo {author} {\bibfnamefont {S.}~\bibnamefont {Puri}},\ and\ \bibinfo {author} {\bibfnamefont {J.~D.}\ \bibnamefont {Thompson}},\ }\bibfield  {title} {\bibinfo {title} {High-fidelity gates and mid-circuit erasure conversion in an atomic qubit},\ }\href {https://doi.org/10.1038/s41586-023-06438-1} {\bibfield  {journal} {\bibinfo  {journal} {Nature}\ }\textbf {\bibinfo {volume} {622}},\ \bibinfo {pages} {279} (\bibinfo {year} {2023})}\BibitemShut {NoStop}%
\bibitem [{\citenamefont {Scholl}\ \emph {et~al.}(2023)\citenamefont {Scholl}, \citenamefont {Shaw}, \citenamefont {Tsai}, \citenamefont {Finkelstein}, \citenamefont {Choi},\ and\ \citenamefont {Endres}}]{Scholl2023eci}%
  \BibitemOpen
  \bibfield  {author} {\bibinfo {author} {\bibfnamefont {P.}~\bibnamefont {Scholl}}, \bibinfo {author} {\bibfnamefont {A.~L.}\ \bibnamefont {Shaw}}, \bibinfo {author} {\bibfnamefont {R.~B.-S.}\ \bibnamefont {Tsai}}, \bibinfo {author} {\bibfnamefont {R.}~\bibnamefont {Finkelstein}}, \bibinfo {author} {\bibfnamefont {J.}~\bibnamefont {Choi}},\ and\ \bibinfo {author} {\bibfnamefont {M.}~\bibnamefont {Endres}},\ }\bibfield  {title} {\bibinfo {title} {Erasure conversion in a high-fidelity {Rydberg} quantum simulator},\ }\href {https://doi.org/https://doi.org/10.1038/s41586-023-06516-4} {\bibfield  {journal} {\bibinfo  {journal} {Nature}\ }\textbf {\bibinfo {volume} {622}},\ \bibinfo {pages} {273} (\bibinfo {year} {2023})}\BibitemShut {NoStop}%
\bibitem [{\citenamefont {Muzi~Falconi}\ \emph {et~al.}(2025)\citenamefont {Muzi~Falconi}, \citenamefont {Panza}, \citenamefont {Sbernardori}, \citenamefont {Forti}, \citenamefont {Klemt}, \citenamefont {Abdel~Karim}, \citenamefont {Marinelli},\ and\ \citenamefont {Scazza}}]{MuziFalconi2025msh}%
  \BibitemOpen
  \bibfield  {author} {\bibinfo {author} {\bibfnamefont {A.}~\bibnamefont {Muzi~Falconi}}, \bibinfo {author} {\bibfnamefont {R.}~\bibnamefont {Panza}}, \bibinfo {author} {\bibfnamefont {S.}~\bibnamefont {Sbernardori}}, \bibinfo {author} {\bibfnamefont {R.}~\bibnamefont {Forti}}, \bibinfo {author} {\bibfnamefont {R.}~\bibnamefont {Klemt}}, \bibinfo {author} {\bibfnamefont {O.}~\bibnamefont {Abdel~Karim}}, \bibinfo {author} {\bibfnamefont {M.}~\bibnamefont {Marinelli}},\ and\ \bibinfo {author} {\bibfnamefont {F.}~\bibnamefont {Scazza}},\ }\bibfield  {title} {\bibinfo {title} {{Microsecond-Scale High-Survival and Number-Resolved Detection of Ytterbium Atom Arrays}},\ }\href {https://doi.org/10.1103/n3bg-7yw7} {\bibfield  {journal} {\bibinfo  {journal} {Phys. Rev. Lett.}\ }\textbf {\bibinfo {volume} {135}},\ \bibinfo {pages} {203402} (\bibinfo {year} {2025})}\BibitemShut {NoStop}%
\bibitem [{\citenamefont {Plassmann}\ \emph {et~al.}(2026)\citenamefont {Plassmann}, \citenamefont {Schaefer}, \citenamefont {Menashes},\ and\ \citenamefont {Salomon}}]{Plassmann2026rsr}%
  \BibitemOpen
  \bibfield  {author} {\bibinfo {author} {\bibfnamefont {T.}~\bibnamefont {Plassmann}}, \bibinfo {author} {\bibfnamefont {L.}~\bibnamefont {Schaefer}}, \bibinfo {author} {\bibfnamefont {M.}~\bibnamefont {Menashes}},\ and\ \bibinfo {author} {\bibfnamefont {G.}~\bibnamefont {Salomon}},\ }\href@noop {} {\bibinfo {title} {Rapid state-resolved single-atom imaging of alkaline-earth fermions}} (\bibinfo {year} {2026}),\ \Eprint {https://arxiv.org/abs/2602.19876} {arXiv:2602.19876 [cond-mat.quant-gas]} \BibitemShut {NoStop}%
\bibitem [{\citenamefont {Sohmen}\ \emph {et~al.}(2023)\citenamefont {Sohmen}, \citenamefont {Mark}, \citenamefont {Greiner},\ and\ \citenamefont {Ferlaino}}]{Sohmen2023asi}%
  \BibitemOpen
  \bibfield  {author} {\bibinfo {author} {\bibfnamefont {M.}~\bibnamefont {Sohmen}}, \bibinfo {author} {\bibfnamefont {M.~J.}\ \bibnamefont {Mark}}, \bibinfo {author} {\bibfnamefont {M.}~\bibnamefont {Greiner}},\ and\ \bibinfo {author} {\bibfnamefont {F.}~\bibnamefont {Ferlaino}},\ }\bibfield  {title} {\bibinfo {title} {A ship-in-a-bottle quantum gas microscope setup for magnetic mixtures},\ }\href {https://doi.org/10.21468/SciPostPhys.15.5.182} {\bibfield  {journal} {\bibinfo  {journal} {SciPost Phys.}\ }\textbf {\bibinfo {volume} {15}},\ \bibinfo {pages} {182} (\bibinfo {year} {2023})}\BibitemShut {NoStop}%
\bibitem [{\citenamefont {Asteria}\ \emph {et~al.}(2021)\citenamefont {Asteria}, \citenamefont {Zahn}, \citenamefont {Kosch}, \citenamefont {Sengstock},\ and\ \citenamefont {Weitenberg}}]{Asteria2021qgm}%
  \BibitemOpen
  \bibfield  {author} {\bibinfo {author} {\bibfnamefont {L.}~\bibnamefont {Asteria}}, \bibinfo {author} {\bibfnamefont {H.~P.}\ \bibnamefont {Zahn}}, \bibinfo {author} {\bibfnamefont {M.~N.}\ \bibnamefont {Kosch}}, \bibinfo {author} {\bibfnamefont {K.}~\bibnamefont {Sengstock}},\ and\ \bibinfo {author} {\bibfnamefont {C.}~\bibnamefont {Weitenberg}},\ }\bibfield  {title} {\bibinfo {title} {Quantum gas magnifier for sub-lattice-resolved imaging of {3D} quantum systems},\ }\href {https://doi.org/https://doi.org/10.1038/s41586-021-04011-2} {\bibfield  {journal} {\bibinfo  {journal} {Nature}\ }\textbf {\bibinfo {volume} {599}},\ \bibinfo {pages} {571–575} (\bibinfo {year} {2021})}\BibitemShut {NoStop}%
\bibitem [{\citenamefont {Murthy}\ \emph {et~al.}(2014)\citenamefont {Murthy}, \citenamefont {Kedar}, \citenamefont {Lompe}, \citenamefont {Neidig}, \citenamefont {Ries}, \citenamefont {Wenz}, \citenamefont {Z\"urn},\ and\ \citenamefont {Jochim}}]{Murthy2014mvf}%
  \BibitemOpen
  \bibfield  {author} {\bibinfo {author} {\bibfnamefont {P.~A.}\ \bibnamefont {Murthy}}, \bibinfo {author} {\bibfnamefont {D.}~\bibnamefont {Kedar}}, \bibinfo {author} {\bibfnamefont {T.}~\bibnamefont {Lompe}}, \bibinfo {author} {\bibfnamefont {M.}~\bibnamefont {Neidig}}, \bibinfo {author} {\bibfnamefont {M.~G.}\ \bibnamefont {Ries}}, \bibinfo {author} {\bibfnamefont {A.~N.}\ \bibnamefont {Wenz}}, \bibinfo {author} {\bibfnamefont {G.}~\bibnamefont {Z\"urn}},\ and\ \bibinfo {author} {\bibfnamefont {S.}~\bibnamefont {Jochim}},\ }\bibfield  {title} {\bibinfo {title} {{Matter-wave Fourier optics with a strongly interacting two-dimensional Fermi gas}},\ }\href {https://doi.org/10.1103/PhysRevA.90.043611} {\bibfield  {journal} {\bibinfo  {journal} {Phys. Rev. A}\ }\textbf {\bibinfo {volume} {90}},\ \bibinfo {pages} {043611} (\bibinfo {year} {2014})}\BibitemShut {NoStop}%
\bibitem [{\citenamefont {Anich}\ \emph {et~al.}(2024)\citenamefont {Anich}, \citenamefont {H\"ollrigl}, \citenamefont {Kreyer}, \citenamefont {Grimm},\ and\ \citenamefont {Kirilov}}]{Anich2024cco}%
  \BibitemOpen
  \bibfield  {author} {\bibinfo {author} {\bibfnamefont {G.}~\bibnamefont {Anich}}, \bibinfo {author} {\bibfnamefont {N.}~\bibnamefont {H\"ollrigl}}, \bibinfo {author} {\bibfnamefont {M.}~\bibnamefont {Kreyer}}, \bibinfo {author} {\bibfnamefont {R.}~\bibnamefont {Grimm}},\ and\ \bibinfo {author} {\bibfnamefont {E.}~\bibnamefont {Kirilov}},\ }\bibfield  {title} {\bibinfo {title} {Comprehensive characterization of an apparatus for cold electromagnetic dysprosium dipoles},\ }\href {https://doi.org/10.1103/PhysRevA.110.023311} {\bibfield  {journal} {\bibinfo  {journal} {Phys. Rev. A}\ }\textbf {\bibinfo {volume} {110}},\ \bibinfo {pages} {023311} (\bibinfo {year} {2024})}\BibitemShut {NoStop}%
\bibitem [{\citenamefont {Baier}\ \emph {et~al.}(2016)\citenamefont {Baier}, \citenamefont {Mark}, \citenamefont {Petter}, \citenamefont {Aikawa}, \citenamefont {Chomaz}, \citenamefont {Cai}, \citenamefont {Baranov}, \citenamefont {Zoller},\ and\ \citenamefont {Ferlaino}}]{Baier2016ebh}%
  \BibitemOpen
  \bibfield  {author} {\bibinfo {author} {\bibfnamefont {S.}~\bibnamefont {Baier}}, \bibinfo {author} {\bibfnamefont {M.~J.}\ \bibnamefont {Mark}}, \bibinfo {author} {\bibfnamefont {D.}~\bibnamefont {Petter}}, \bibinfo {author} {\bibfnamefont {K.}~\bibnamefont {Aikawa}}, \bibinfo {author} {\bibfnamefont {L.}~\bibnamefont {Chomaz}}, \bibinfo {author} {\bibfnamefont {Z.}~\bibnamefont {Cai}}, \bibinfo {author} {\bibfnamefont {M.}~\bibnamefont {Baranov}}, \bibinfo {author} {\bibfnamefont {P.}~\bibnamefont {Zoller}},\ and\ \bibinfo {author} {\bibfnamefont {F.}~\bibnamefont {Ferlaino}},\ }\bibfield  {title} {\bibinfo {title} {Extended {Bose-Hubbard} models with ultracold magnetic atoms},\ }\href {https://doi.org/10.1126/science.aac9812} {\bibfield  {journal} {\bibinfo  {journal} {Science}\ }\textbf {\bibinfo {volume} {352}},\ \bibinfo {pages} {201} (\bibinfo {year} {2016})}\BibitemShut {NoStop}%
\bibitem [{\citenamefont {Impertro}\ \emph {et~al.}(2023)\citenamefont {Impertro}, \citenamefont {Wienand}, \citenamefont {Häfele}, \citenamefont {von Raven}, \citenamefont {Hubele}, \citenamefont {Klostermann}, \citenamefont {Cabrera}, \citenamefont {Bloch},\ and\ \citenamefont {Aidelsburger}}]{Impertro2023aud}%
  \BibitemOpen
  \bibfield  {author} {\bibinfo {author} {\bibfnamefont {A.}~\bibnamefont {Impertro}}, \bibinfo {author} {\bibfnamefont {J.~F.}\ \bibnamefont {Wienand}}, \bibinfo {author} {\bibfnamefont {S.}~\bibnamefont {Häfele}}, \bibinfo {author} {\bibfnamefont {H.}~\bibnamefont {von Raven}}, \bibinfo {author} {\bibfnamefont {S.}~\bibnamefont {Hubele}}, \bibinfo {author} {\bibfnamefont {T.}~\bibnamefont {Klostermann}}, \bibinfo {author} {\bibfnamefont {C.~R.}\ \bibnamefont {Cabrera}}, \bibinfo {author} {\bibfnamefont {I.}~\bibnamefont {Bloch}},\ and\ \bibinfo {author} {\bibfnamefont {M.}~\bibnamefont {Aidelsburger}},\ }\bibfield  {title} {\bibinfo {title} {An unsupervised deep learning algorithm for single-site reconstruction in quantum gas microscopes.},\ }\href {https://doi.org/https://doi.org/10.1038/s42005-023-01287-w} {\bibfield  {journal} {\bibinfo  {journal} {Communications Physics}\ }\textbf {\bibinfo {volume} {6}},\ \bibinfo {pages} {166} (\bibinfo {year} {2023})}\BibitemShut {NoStop}%
\bibitem [{\citenamefont {La~Rooij}\ \emph {et~al.}(2023)\citenamefont {La~Rooij}, \citenamefont {Ulm}, \citenamefont {Haller},\ and\ \citenamefont {Kuhr}}]{LaRooij2023acs}%
  \BibitemOpen
  \bibfield  {author} {\bibinfo {author} {\bibfnamefont {A.}~\bibnamefont {La~Rooij}}, \bibinfo {author} {\bibfnamefont {C.}~\bibnamefont {Ulm}}, \bibinfo {author} {\bibfnamefont {E.}~\bibnamefont {Haller}},\ and\ \bibinfo {author} {\bibfnamefont {S.}~\bibnamefont {Kuhr}},\ }\bibfield  {title} {\bibinfo {title} {A comparative study of deconvolution techniques for quantum-gas microscope images},\ }\href {https://doi.org/10.1088/1367-2630/aced65} {\bibfield  {journal} {\bibinfo  {journal} {New Journal of Physics}\ }\textbf {\bibinfo {volume} {25}},\ \bibinfo {pages} {083036} (\bibinfo {year} {2023})}\BibitemShut {NoStop}%
\bibitem [{\citenamefont {Stetson}(1994)}]{Stetson1994tco}%
  \BibitemOpen
  \bibfield  {author} {\bibinfo {author} {\bibfnamefont {P.~B.}\ \bibnamefont {Stetson}},\ }\bibfield  {title} {\bibinfo {title} {{The Center of the Core-Cusp Globular Cluster {M15}: {CFHT} and {HST} Observations, {ALLFRAME} Reductions}},\ }\href {https://doi.org/10.1086/133378} {\bibfield  {journal} {\bibinfo  {journal} {Publications of the Astronomical Society of the Pacific}\ }\textbf {\bibinfo {volume} {106}},\ \bibinfo {pages} {250} (\bibinfo {year} {1994})}\BibitemShut {NoStop}%
\bibitem [{\citenamefont {Lelek}\ \emph {et~al.}(2021)\citenamefont {Lelek}, \citenamefont {Gyparaki}, \citenamefont {Beliu}, \citenamefont {Schueder}, \citenamefont {Griffi{\'e}}, \citenamefont {Manley}, \citenamefont {Jungmann}, \citenamefont {Sauer}, \citenamefont {Lakadamyali},\ and\ \citenamefont {Zimmer}}]{Lelek2021sml}%
  \BibitemOpen
  \bibfield  {author} {\bibinfo {author} {\bibfnamefont {M.}~\bibnamefont {Lelek}}, \bibinfo {author} {\bibfnamefont {M.~T.}\ \bibnamefont {Gyparaki}}, \bibinfo {author} {\bibfnamefont {G.}~\bibnamefont {Beliu}}, \bibinfo {author} {\bibfnamefont {F.}~\bibnamefont {Schueder}}, \bibinfo {author} {\bibfnamefont {J.}~\bibnamefont {Griffi{\'e}}}, \bibinfo {author} {\bibfnamefont {S.}~\bibnamefont {Manley}}, \bibinfo {author} {\bibfnamefont {R.}~\bibnamefont {Jungmann}}, \bibinfo {author} {\bibfnamefont {M.}~\bibnamefont {Sauer}}, \bibinfo {author} {\bibfnamefont {M.}~\bibnamefont {Lakadamyali}},\ and\ \bibinfo {author} {\bibfnamefont {C.}~\bibnamefont {Zimmer}},\ }\bibfield  {title} {\bibinfo {title} {Single-molecule localization microscopy},\ }\href {https://doi.org/10.1038/s43586-021-00038-x} {\bibfield  {journal} {\bibinfo  {journal} {Nature Reviews Methods Primers}\ }\textbf {\bibinfo {volume} {1}},\ \bibinfo {pages} {39} (\bibinfo {year} {2021})}\BibitemShut {NoStop}%
\bibitem [{\citenamefont {Bloch}\ \emph {et~al.}(2008)\citenamefont {Bloch}, \citenamefont {Dalibard},\ and\ \citenamefont {Zwerger}}]{Bloch2008mbp}%
  \BibitemOpen
  \bibfield  {author} {\bibinfo {author} {\bibfnamefont {I.}~\bibnamefont {Bloch}}, \bibinfo {author} {\bibfnamefont {J.}~\bibnamefont {Dalibard}},\ and\ \bibinfo {author} {\bibfnamefont {W.}~\bibnamefont {Zwerger}},\ }\bibfield  {title} {\bibinfo {title} {{Many-body physics with ultracold gases}},\ }\href {https://doi.org/10.1103/RevModPhys.80.885} {\bibfield  {journal} {\bibinfo  {journal} {Rev. Mod. Phys.}\ }\textbf {\bibinfo {volume} {80}},\ \bibinfo {pages} {885} (\bibinfo {year} {2008})}\BibitemShut {NoStop}%
\bibitem [{\citenamefont {Ketterle}\ and\ \citenamefont {Zwierlein}(2008)}]{Ketterle2008mpa}%
  \BibitemOpen
  \bibfield  {author} {\bibinfo {author} {\bibfnamefont {W.}~\bibnamefont {Ketterle}}\ and\ \bibinfo {author} {\bibfnamefont {M.~W.}\ \bibnamefont {Zwierlein}},\ }\bibfield  {title} {\bibinfo {title} {Making, probing and understanding ultracold fermi gases},\ }\href {https://doi.org/10.1393/ncr/i2008-10033-1} {\bibfield  {journal} {\bibinfo  {journal} {La Rivista del Nuovo Cimento}\ }\textbf {\bibinfo {volume} {31}},\ \bibinfo {pages} {247} (\bibinfo {year} {2008})}\BibitemShut {NoStop}%
\bibitem [{\citenamefont {Hadzibabic}\ and\ \citenamefont {Dalibard}(2011)}]{Hadzibabic2011}%
  \BibitemOpen
  \bibfield  {author} {\bibinfo {author} {\bibfnamefont {Z.}~\bibnamefont {Hadzibabic}}\ and\ \bibinfo {author} {\bibfnamefont {J.}~\bibnamefont {Dalibard}},\ }\bibfield  {title} {\bibinfo {title} {Two-dimensional {Bose} fluids: An atomic physics perspective},\ }\href {https://doi.org/10.1393/ncr/i2011-10066-3} {\bibfield  {journal} {\bibinfo  {journal} {La Rivista del Nuovo Cimento}\ }\textbf {\bibinfo {volume} {34}},\ \bibinfo {pages} {389} (\bibinfo {year} {2011})}\BibitemShut {NoStop}%
\bibitem [{\citenamefont {Dalibard}\ \emph {et~al.}(2011)\citenamefont {Dalibard}, \citenamefont {Gerbier}, \citenamefont {Juzeliunas},\ and\ \citenamefont {\"Ohberg}}]{Dalibard2011agp}%
  \BibitemOpen
  \bibfield  {author} {\bibinfo {author} {\bibfnamefont {J.}~\bibnamefont {Dalibard}}, \bibinfo {author} {\bibfnamefont {F.}~\bibnamefont {Gerbier}}, \bibinfo {author} {\bibfnamefont {G.}~\bibnamefont {Juzeliunas}},\ and\ \bibinfo {author} {\bibfnamefont {P.}~\bibnamefont {\"Ohberg}},\ }\bibfield  {title} {\bibinfo {title} {{Artificial gauge potentials for neutral atoms}},\ }\href {https://doi.org/10.1103/RevModPhys.83.1523} {\bibfield  {journal} {\bibinfo  {journal} {Rev. Mod. Phys.}\ }\textbf {\bibinfo {volume} {83}},\ \bibinfo {pages} {1523} (\bibinfo {year} {2011})}\BibitemShut {NoStop}%
\bibitem [{\citenamefont {Madeira}\ \emph {et~al.}(2020)\citenamefont {Madeira}, \citenamefont {Cidrim}, \citenamefont {Hemmerling}, \citenamefont {Caracanhas}, \citenamefont {dos Santos},\ and\ \citenamefont {Bagnato}}]{Madeira2020qtb}%
  \BibitemOpen
  \bibfield  {author} {\bibinfo {author} {\bibfnamefont {L.}~\bibnamefont {Madeira}}, \bibinfo {author} {\bibfnamefont {A.}~\bibnamefont {Cidrim}}, \bibinfo {author} {\bibfnamefont {M.}~\bibnamefont {Hemmerling}}, \bibinfo {author} {\bibfnamefont {M.~A.}\ \bibnamefont {Caracanhas}}, \bibinfo {author} {\bibfnamefont {F.~E.~A.}\ \bibnamefont {dos Santos}},\ and\ \bibinfo {author} {\bibfnamefont {V.~S.}\ \bibnamefont {Bagnato}},\ }\bibfield  {title} {\bibinfo {title} {Quantum turbulence in bose–einstein condensates: Present status and new challenges ahead},\ }\href {https://doi.org/10.1116/5.0016751} {\bibfield  {journal} {\bibinfo  {journal} {AVS Quantum Science}\ }\textbf {\bibinfo {volume} {2}},\ \bibinfo {pages} {035901} (\bibinfo {year} {2020})}\BibitemShut {NoStop}%
\bibitem [{\citenamefont {Ilzh\"ofer}\ \emph {et~al.}(2018)\citenamefont {Ilzh\"ofer}, \citenamefont {Durastante}, \citenamefont {Patscheider}, \citenamefont {Trautmann}, \citenamefont {Mark},\ and\ \citenamefont {Ferlaino}}]{Ilzhoefer2018tsf}%
  \BibitemOpen
  \bibfield  {author} {\bibinfo {author} {\bibfnamefont {P.}~\bibnamefont {Ilzh\"ofer}}, \bibinfo {author} {\bibfnamefont {G.}~\bibnamefont {Durastante}}, \bibinfo {author} {\bibfnamefont {A.}~\bibnamefont {Patscheider}}, \bibinfo {author} {\bibfnamefont {A.}~\bibnamefont {Trautmann}}, \bibinfo {author} {\bibfnamefont {M.~J.}\ \bibnamefont {Mark}},\ and\ \bibinfo {author} {\bibfnamefont {F.}~\bibnamefont {Ferlaino}},\ }\bibfield  {title} {\bibinfo {title} {{Two-species five-beam magneto-optical trap for erbium and dysprosium}},\ }\href {https://doi.org/10.1103/PhysRevA.97.023633} {\bibfield  {journal} {\bibinfo  {journal} {Phys. Rev. A}\ }\textbf {\bibinfo {volume} {97}},\ \bibinfo {pages} {023633} (\bibinfo {year} {2018})}\BibitemShut {NoStop}%
\bibitem [{\citenamefont {Wenzel}(2015)}]{Wenzel2015adq}%
  \BibitemOpen
  \bibfield  {author} {\bibinfo {author} {\bibfnamefont {M.}~\bibnamefont {Wenzel}},\ }\emph {\bibinfo {title} {A dysprosium quantum gas in highly controllable optical traps}},\ \href {https://www.pi5.uni-stuttgart.de/documents/abgeschlossene-arbeiten/2015-Wenzel-Matthias-A-dysprosium-quantum-gas-in-highly-controllable-optical-traps-MSC.pdf} {\bibinfo {type} {Master's thesis}},\ \bibinfo  {school} {Universit{\"a}t Stuttgart} (\bibinfo {year} {2015})\BibitemShut {NoStop}%
\bibitem [{\citenamefont {Léonard}\ \emph {et~al.}(2014)\citenamefont {Léonard}, \citenamefont {Lee}, \citenamefont {Morales}, \citenamefont {Karg}, \citenamefont {Esslinger},\ and\ \citenamefont {Donner}}]{Leonard2014oto}%
  \BibitemOpen
  \bibfield  {author} {\bibinfo {author} {\bibfnamefont {J.}~\bibnamefont {Léonard}}, \bibinfo {author} {\bibfnamefont {M.}~\bibnamefont {Lee}}, \bibinfo {author} {\bibfnamefont {A.}~\bibnamefont {Morales}}, \bibinfo {author} {\bibfnamefont {T.~M.}\ \bibnamefont {Karg}}, \bibinfo {author} {\bibfnamefont {T.}~\bibnamefont {Esslinger}},\ and\ \bibinfo {author} {\bibfnamefont {T.}~\bibnamefont {Donner}},\ }\bibfield  {title} {\bibinfo {title} {Optical transport and manipulation of an ultracold atomic cloud using focus-tunable lenses},\ }\href {https://doi.org/10.1088/1367-2630/16/9/093028} {\bibfield  {journal} {\bibinfo  {journal} {New Journal of Physics}\ }\textbf {\bibinfo {volume} {16}},\ \bibinfo {pages} {093028} (\bibinfo {year} {2014})}\BibitemShut {NoStop}%
\bibitem [{\citenamefont {Lavigne}\ \emph {et~al.}(2022)\citenamefont {Lavigne}, \citenamefont {Groh},\ and\ \citenamefont {Stellmer}}]{lavigne2022mot}%
  \BibitemOpen
  \bibfield  {author} {\bibinfo {author} {\bibfnamefont {Q.}~\bibnamefont {Lavigne}}, \bibinfo {author} {\bibfnamefont {T.}~\bibnamefont {Groh}},\ and\ \bibinfo {author} {\bibfnamefont {S.}~\bibnamefont {Stellmer}},\ }\bibfield  {title} {\bibinfo {title} {Magneto-optical trapping of mercury at high phase-space density},\ }\href {https://doi.org/10.1103/PhysRevA.105.033106} {\bibfield  {journal} {\bibinfo  {journal} {Physical Review A}\ }\textbf {\bibinfo {volume} {105}},\ \bibinfo {pages} {033106} (\bibinfo {year} {2022})}\BibitemShut {NoStop}%
\end{thebibliography}
\end{document}